\documentclass[twocolumn]{aastex63}

\shorttitle{AGN Feedback in SDSS-IV MaNGA}
\shortauthors{Lammers et al.}

\graphicspath{{./}{}}
\usepackage{amsmath}

\begin{document}

\title{AGN Feedback in SDSS-IV MaNGA: AGNs have Suppressed Central Star Formation Rates}


\author[0000-0001-9985-0643]{Caleb Lammers}
\affiliation{Dunlap Institute for Astronomy \& Astrophysics, University of Toronto, 50 St.\ George Street, Toronto, ON M5S 3H4, Canada}
\affiliation{Department of Physics, University of Toronto, 60 St.\ George Street, Toronto, ON M5S 1A7, Canada}

\author[0000-0001-9298-3523]{Kartheik G.\ Iyer}
\altaffiliation{Hubble Fellow}
\affiliation{Dunlap Institute for Astronomy \& Astrophysics, University of Toronto, 50 St.\ George Street, Toronto, ON M5S 3H4, Canada}
\affiliation{Columbia Astrophysics Laboratory, Columbia University, 550 West 120th Street, New York, NY 10027, USA}

\author[0000-0002-9790-6313]{Hector Ibarra-Medel}
\affiliation{Escuela Superior de F\'{\i}sica y Matem\'aticas, Instituto Polit\'ecnico Nacional, U.P. Adolfo L\'opez Mateos, C.P. 07738, Ciudad de M\'exico, M\'exico}
\affiliation{Instituto de Astronom\'ia y Ciencias Planetarias, Universidad de Atacama, Copayapu 485, Copiap\'o, Chile}

\author[0000-0003-4196-0617]{Camilla Pacifici}
\affiliation{Space Telescope Science Institute, 3700 San Martin Drive, Baltimore, MD 21218, USA}

\author[0000-0001-6444-9307]{Sebastián F.\ Sánchez}
\affiliation{Instituto de Astronomía, Universidad Nacional Autónoma de México, AP 70-264, CDMX 04510, México}

\author[0000-0002-8224-4505]{Sandro Tacchella}
\affiliation{Kavli Institute for Cosmology, University of Cambridge, Madingley Road, Cambridge, CB3 0HA, UK}
\affiliation{Cavendish Laboratory, University of Cambridge, 19 JJ Thomson Avenue, Cambridge, CB3 0HE, UK}

\author{Joanna Woo}
\affiliation{Department of Physics, Simon Fraser University, 8888 University Drive, Burnaby BC V5A 1S6, Canada}

\begin{abstract}

Despite the importance of feedback from active galactic nuclei (AGNs) in models of galaxy evolution, observational constraints on the influence of AGN feedback on star formation remain weak. To this end, we have compared the star formation trends of 279 low-redshift AGN galaxies with 558 inactive control galaxies using integral field unit spectroscopy from the SDSS-IV MaNGA survey. With a Gaussian process-based methodology, we reconstruct nonparametric star formation histories in spatially resolved spaxels covering the face of each galaxy. Based on galaxy-wide star formation rates (SFRs) alone, we find no obvious signatures of AGN feedback. However, the AGN galaxies have significantly suppressed central (kiloparsec-scale) SFRs, lying up to a factor of $2$ below those of the control galaxies, providing direct observational evidence of AGN feedback suppressing star formation. The suppression of central SFRs in the AGN galaxies began in the central regions ${\sim}\,6$\,Gyr ago (redshift $z\,{\sim}\,0.7$), taking place over a few gigayears. A small subset of the AGN galaxies were rapidly driven to quiescence shortly before being observed (in the last $500$\,Myr), potentially indicating instances of AGN-driven feedback. More frequently, however, star formation continues in the AGN galaxies, with suppression primarily in the central regions. This is suggestive of a picture in which integrated (Gyr-timescale) AGN feedback can significantly affect central star formation, but may be inefficient in driving galaxy-wide quenching in low-redshift galaxies, instead leaving them in the green valley.

\end{abstract}


\section{Introduction}
\label{sec:intro}

It is now widely believed that the evolution of galaxies is tied to the growth of their central supermassive black holes (SMBHs). In support of this, a number of connections between the properties of SMBHs (such as mass) and the properties of their host galaxies (such as stellar mass, bulge luminosity, and concentration) have been established observationally \citep[e.g.,][]{MagorrianAGN, FerrareseAGN, MarconiAGN, KormendyAGN}. During periods of accretion, SMBHs can release substantial amounts of energy back into their host galaxies --- this is termed ``active galactic nucleus'' (AGN) feedback. In fact, it has been demonstrated theoretically that luminous AGNs can release enough energy to remove all gas from their host galaxies \citep{SilkReesAGN, KingAGNenergy, MurrayAGNenergy}. The SMBH-galaxy connection can be understood, at least in part, if feedback from AGNs regulates the growth of galaxies. However, the extent to which AGN feedback actually affects the evolution of galaxies remains somewhat unclear.

Observations indicate that luminous AGNs, found primarily at high redshifts, can drive rapid outflows that significantly impact galaxy-wide gas content \citep[e.g.,][]{SturmAGNoutflow, MaiolinoAGNoutflow, CiconeAGNoutflow, KakkadAGNoutflow, ForsterSchreiberoutflow}. Furthermore, on the theoretical end, numerical simulations indicate that AGN feedback in luminous AGNs plays an important role in producing quiescent, red galaxies \citep{DiMatteoAGNsim, CattaneoAGNsim, CrotonAGNsim, PuchweinAGNsim, VogelsbergerAGNsim, WeinbergerAGNsim}. However, far more galaxies (particularly at low redshifts) host low-luminosity, Seyfert-type AGNs \citep{Maiolino&RiekeSeyferts, HoSeyfertsreview}. Despite significant interest, the extent to which typical AGNs influence the evolution of their host galaxies is not yet understood. A number of studies have found that star formation in typical AGN host galaxies is consistent with that of non-AGN galaxies at the same evolutionary stage \citep[e.g.,][]{BongiornoAGNSFMS, ChangAGNSFMS, SuhAGNSFMS, first62AGNIV}. On the other hand, many studies have found that AGN galaxies have preferentially lower star formation rates \citep[SFRs; e.g.,][]{SchawinskiAGNhighSM, SilvermanAGNs, MullaneyAGNSFMS, SanchezAGNSFMS, LacerdaAGNSFMS} or preferentially higher SFRs \citep[e.g.,][]{KossAGNSFMS, SantiniAGNSFMS, EllisonAGNSFMS, WooAGNSFMS}, potentially indicating the influence of AGN feedback. Inevitably, studies of AGN host galaxies are limited by the quality of data (particularly in the spatial dimension) and selection effects introduced by the choice of AGN selection criteria, both of which can potentially be improved with integral field unit (IFU) spectroscopy.

IFU spectroscopy galaxy surveys, including the Calar Alto Legacy Integral Field Area (CALIFA; \citealt{SanchezCALIFA}) survey, the Sydney-AAO Multi-object Integral field (SAMI; \citealt{CroomSAMI}) survey, and the Mapping Nearby Galaxies at Apache Point Observatory (MaNGA; \citealt{BundyMaNGA}) survey, have revolutionized our understanding of spatial trends in galaxy evolution. In these surveys, spectroscopy is carried out in each spatially resolved pixel, thereby transforming the pixels in a conventional image to spectral pixels (termed ``spaxels''). The spatial dimension provided by IFU data has enabled studies on the spatial trends in local galaxies, revealing that massive galaxies have centrally suppressed SFRs, both at high redshifts \citep{TacchellahighzIFU1, TacchellahighzIFU2} and low redshifts \citep{BelfioreradialSFR, EllisonradialSFR, Bluckquenching}. This hints at the importance of inside out quenching mechanisms, such as AGN feedback, morphological quenching, and stellar feedback. The ability to explore spatial trends is particularly relevant for studies of AGN feedback, for which simulations suggest that the central regions of galaxies may be affected most significantly \citep{HopkinsAGNsim, ApplebyAGNsim, TorreyFIREAGN, NelsonTNGquenching}. In addition to enabling the spatially resolved study of galaxies, IFU data has been leveraged to identify AGNs that may be missed with conventional single-fiber spectra, potentially allowing for more complete samples of AGN galaxies \citep{WylezalekhiddenAGNs, Wylezalekcrit}. This is particularly relevant for AGNs hidden by dust or contained within galaxies dominated by star formation, as well as AGNs that have recently turned off. In this work, we use MaNGA data to study the influence of AGN feedback on host galaxy star formation, taking advantage of MaNGA-specific AGN selection criteria \citep{Wylezalekcrit}.

Extracting physical information from the spectra of galaxies is a classical problem in astronomy, facilitated with modern, high-quality spectroscopy and sophisticated statistical techniques. Traditionally, galaxy SFRs were estimated based on H$\alpha$ luminosity \citep{KennicuttSFR1, KennicuttSFR2, KennicuttSFR3}. Unfortunately, AGNs can also contribute to H$\alpha$ emission, potentially biasing the SFRs measured for AGN galaxies. Although the bias in galaxy-wide SFR is small for typical Seyfert/LINER AGNs \citep{TorrecillaCALIFAAGN1, TorrecillaCALIFAAGN2, SanchezAGNSFMS}, AGN contamination can dominate H$\alpha$ emission in individual IFU spaxels \citep{first62AGNIV}. Instead, we can recover accurate spaxel/galaxy properties by fitting a model parameterized with physical properties (stellar mass, SFR, stellar ages, etc.) to the observed spectral energy distribution (termed SED fitting). When recovering stellar population information, emission lines are often masked \citep{SanchezPipe3D1, LacerdaPipe3D}, mitigating the influence of AGN contamination. In addition to inferring galaxy/spaxel SFRs, SED fitting can be used to recover star formation histories (SFHs). SFHs provide a record of when galaxies formed their stars, allowing for insights into the physical processes governing star formation, such as inflows of gas, feedback from AGN/supernovae, and mergers \citep[e.g.,][]{Kauffmanbimod, BrinchmannSFMS, SomervilleSEDfitting, AbramsonSEDfitting, BehrooziUNIVERSEMACHINE, IyerSFHs, TacchellaSFHs2}. The simplest SFH reconstruction techniques employ parametric SFH models, such as exponentially declining SFHs, which can introduce biases in the recovered parameters and miss multiple periods of star formation/quenching \citep{Cieslabias, Carnallbias}. Flexible, nonparametric SFHs minimize bias \citep{Iyerdb1, Lejabias}, but require strong constraints on the underlying stellar populations, such as those provided from SED fitting of MaNGA spectra \citep{SanchezPipe3D2, IbarraMedelMaNGAPIPE3D, SanchezMaNGAPIPE3D}.

The goal of this work is to study the influence of AGN feedback on star formation with nonparametric, spatially resolved SFHs reconstructed from MaNGA data. Section~\ref{sec:data} describes the data used in our analysis, and Section~\ref{sec:methods} details the adopted AGN selection criteria and the PIPE3D Gaussian process SFH reconstruction methodology. In Section~\ref{sec:SFtrends} and Section~\ref{sec:qbehavior}, we compare trends in star formation and quenching between the AGN and control samples. Discussion and conclusions are presented in Section~\ref{sec:discussion} and \ref{sec:conclusion}, respectively. Throughout this paper, we adopt a standard $\Lambda$CDM cosmology with $\Omega_{m}\,{=}\,0.3$, $\Omega_{\Lambda}\,{=}\,0.7$, and H$_{0}\,{=}\,70$\,km\,Mpc$^{-1}$\,s$^{-1}$.

\section{Observations and Data}
\label{sec:data}

\subsection{SDSS-IV MaNGA Survey}
\label{sec:MaNGAsurvey}

The Sloan Digital Sky Survey IV (SDSS-IV) MaNGA survey collects optical spectroscopic IFU data for nearby galaxies \citep{BundyMaNGA, BlantonSDSSIV}. In data release 16 (DR16), observations were reported for 4621 unique galaxies in the redshift range $0.01\,{<}\,z\,{<}\,0.15$. The MaNGA survey uses the BOSS spectrograph on the 2.5\,m SDSS telescope at the Apache Point Observatory, which provides a wavelength coverage from 3600\,--\,10300\,\AA\ \citep{SmeeBOSS, DroryMaNGA}. MaNGA observations are made with integral field units that vary from 19 to 127 fibers. Each optical fiber subtends $\sim\,2$\,arcsec on the sky, and a spaxel size of 0.5\,arcsec is reported in the resulting data cubes, corresponding to a physical distance of ${\sim}\,0.1$\,--\,$1$\,kpc over the MaNGA redshift range \citep{LawDRP}. Observations are calibrated, then reduced with the MaNGA Data Reduction Pipeline (DRP; \citealt{LawDRP}) before physical properties are derived with the MaNGA Data Analysis Pipeline (DAP; \citealt{WestfallDAP}). The MaNGA galaxy target selection criteria were developed to maximize the signal-to-noise ratio (S/N), spatial resolution, and sample size, while maintaining a flat $\log($stellar mass$)$ distribution.

\subsection{Galaxy Zoo}
\label{sec:galaxyzoo}

Galaxy Zoo is a citizen science project that classifies the morphology of observed galaxies via survey-style questions about the galaxies' appearances \citep{WillettGZ2, HartGZ}. Data from Galaxy Zoo 2, as well as unpublished data from Galaxy Zoo 4 and Galaxy Zoo 5, provide a large dataset of visual morphological classifications (smooth vs.\ features/disk, edge-on vs.\ face-on, bulge prominence, etc.) of SDSS galaxies.

\section{Sample Selection and SFH Reconstruction}
\label{sec:methods}

\subsection{Summary of MaNGA AGN Samples}
\label{sec:MaNGAAGNs}

Identifying AGN host galaxies is a challenging problem, for which a variety of selection criteria have been developed. To date, there have been several efforts to identify populations of AGN galaxies in MaNGA, beginning with the ancillary AGN program.\footnote{https://www.sdss.org/dr16/manga/manga-target-selection/ancillary-targets/luminous-agn/} The goal of the ancillary AGN program was to observe luminous ($L_{\mathrm{bol}}\,{>}\,10^{43}$\,erg\,s$^{-1}$) type 2 AGNs, in hopes of gaining insight into the properties of their host galaxies and the narrow-line region. In this program, AGN galaxies were selected based on the Swift/BAT X-ray catalog, [O\,III] $\lambda\,5007$ luminosities, and Wide-field Infrared Survey Explorer (WISE) infrared colors, leading the MaNGA survey to observe 24 ancillary AGN galaxies in DR16. In a similar vein, \citet{ComerfordAGNMaNGA} cross-referenced MaNGA galaxies with various AGN catalogs to construct a sample of 406 AGN galaxies in MPL-8 (289 in the publicly available DR16). Specifically, \citet{ComerfordAGNMaNGA} identified MaNGA AGNs using Swift/BAT X-ray detections, WISE colors, NVSS/SWIFT radio observations, and SDSS broad emission lines (the majority of the AGN catalog came from radio observations).

Instead, most MaNGA AGN studies have relied on optical-based selection criteria, namely, the BPT \citep{BPT1981, Kewleystarburst, KewleyBPT} and WHAN \citep{FernandesWHAN} diagrams. \citet{first62AGNI} and subsequent papers \citep{first62AGNII, first62AGNIII, first62AGNIV}, used integrated SDSS-III spectra to identify galaxies that pass both the BPT and WHAN criteria (i.e., galaxies identified as LINER/Seyfert in the BPT diagram and weak AGN/strong AGN in the WHAN diagram), resulting in a sample of 62 MaNGA AGNs. Similarly, \citet{SanchezAGNSFMS} selected AGN galaxies using the central 3\,arsec\,$\times$\,3\,arcsec integrated MaNGA spectra and BPT/WHAN-like criteria (specifically, galaxies identified as LINER/Seyfert with H$\alpha$ equivalent width EW(H$\alpha$)\,$>$\,1.5\,\AA). This selection procedure resulted in a total of 98 MaNGA AGN galaxies.

Instead of relying on the integrated spectra of MaNGA galaxies, AGN host galaxies can also be identified using the spectra of individual spaxels. In particular, \citet{Wylezalekcrit} (hereafter, W18) developed a systematic approach to classify individual spaxels using the BPT and WHAN diagrams. AGN galaxies are then identified based on the fraction of AGN-classified spaxels, H$\alpha$ surface brightness, and H$\alpha$ EW. This led W18 to identify 303 MaNGA AGNs in MPL-5, many of which were missed with conventional BPT/WHAN-based selection criteria, despite significant signatures of AGN activity. Follow-up observations of 10 potential AGNs from their IFU selection criteria, which do not pass integrated BPT/WHAN criteria, revealed that $7$\,--\,$10$ (depending on criteria strictness) are confirmed AGNs \citep{ComerfordAGNfollowup}. The authors conclude that spatially resolved AGN diagnostics may allow for the identification of low-luminosity AGNs, obscured AGNs, and flickering AGNs that are missed by other selection criteria, potentially allowing for a more complete census of optical AGNs.

Selecting a sample of AGN galaxies is particularly important for studies of host galaxy properties --- both because different populations of AGNs may be at different evolutionary stages \citep{HickoxAGNs} and because missed AGNs can end up in the sample of control galaxies, making trends less clear. Of the MaNGA-based studies, \citet{first62AGNIV} found no signs of AGN feedback suppressing star formation in their sample of 62 BPT/WHAN-identified AGNs when compared with a matched control sample of 109 galaxies. Despite similar AGN selection methodology, \citet{SanchezAGNSFMS} concluded that their sample of 98 AGN galaxies was in the process of halting star formation by comparing with 2700 (unmatched) control galaxies. On the other hand, with a different AGN sample, \citet{ComerfordAGNMaNGA} found evidence of radio-mode AGN galaxies quenching star formation, by comparing their 406 AGN galaxies with the main sequence.

\subsection{AGN Sample Selection}
\label{sec:AGNsample}

\begin{figure}
\centering
\includegraphics[width=0.45\textwidth]{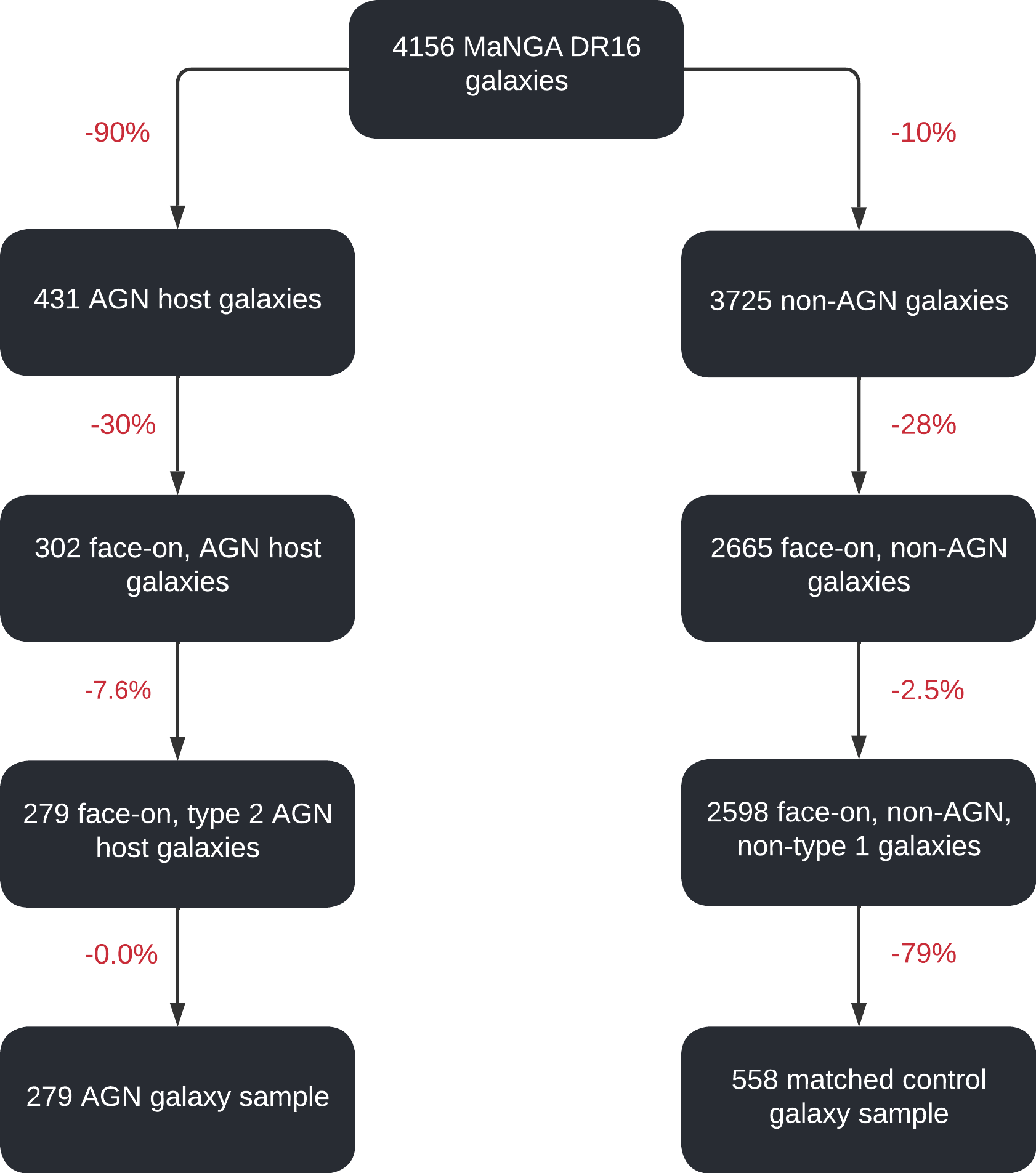}
\caption{Summary of the selection methodology for the sample of 279 AGN galaxies and the sample of 558 stellar mass- and redshift-matched control galaxies. The AGN selection criteria from W18 are applied at step 1.}
\label{fig:selectionmethod}
\end{figure}

We adopt the MaNGA-based, spatially resolved AGN selection criteria from W18 for two primary reasons: sample size and sample completeness. The W18 selection criteria allow us to study a large sample of face-on, type 2 AGNs that are suitable for spatially resolved SED fitting. Additionally, high sample completeness (i.e., few missed AGNs) is crucial for constructing a meaningful matched control sample. It is, however, worth reminding the reader that ``AGNs'' identified based on optical AGN selection criteria are better termed ``AGN candidates'' because several physical processes can mimic optical AGN-like signatures (e.g., shocks and young hot stars). We do not expect a small number of spurious AGNs in the AGN sample to significantly impact results. Indeed, our results are similar (but less statistically significant) if we restrict our analysis to the AGNs that pass integrated BPT/WHAN selection criteria (see Appendix~\ref{sec:selection}).

The AGN selection criteria developed in W18 rely on classifying individual spaxels based on the BPT diagram, H$\alpha$ EW and surface brightness, and the distance from the standard BPT diagnostic lines in line-ratio space. Galaxies are identified as AGN candidates based on thresholds in the calculated parameters and the fraction of spaxels that satisfy various criteria. The full list of parameters and the refined selection criteria can be found in W18. All emission line and EW data used in the selection algorithm are taken from the MaNGA DAP. Before fitting, all spaxels with an r-band S/N\,${<}\,5$ are excluded from the analysis. All remaining spaxels are independently classified according to their positions in the [N\,II]-BPT and [S\,II]-BPT diagrams (as defined in \citealt{KewleyBPT}).

Of the 4156 unique galaxies in MaNGA DR16 without quality flags (i.e., without data collection/analysis issues), we find that the W18 selection criteria identify 431 galaxies as candidate AGN hosts ($\sim\,10$\,\%), similar to the proportion of galaxies identified in the earlier MPL-8 data release by W18. For our analysis, we introduce two additional quality cuts to the AGN sample: excluding edge-on galaxies and excluding type 1 AGNs with potential SED contamination (see below for details).

To fully leverage the spatially resolved data provided by MaNGA, we limit our analysis to face-on galaxies, which allows for the study of radial trends out to ${\sim}\,2.5\,R_e$ while minimizing the deleterious effects of extinction. To remove edge-on galaxies from the AGN sample, we turn to the morphological classifications provided by Galaxy Zoo. For galaxies with three or more votes, those with an ``edge-on appearance'' debiased vote fraction greater than 50\,\% (or a ``contaminating star/artifact'' debiased vote fraction greater than 50\,\%) are excluded from the AGN sample. In the sample of 431 AGN host galaxies, 74 were classified as edge-on, four were classified as containing a contaminating star/artifact, and 51 had insufficient galaxy zoo data, leaving 302 face-on AGN host galaxies.

Performing SED fitting on AGN host galaxies is challenging due to the difficulty of disentangling stellar emission from AGN contamination \citep[e.g.,][]{YangAGNSEDfitting}. This is a particularly pertinent concern for IFU surveys, wherein the spectra of some spaxels may be dominated by AGN contamination. Furthermore, in the MaNGA wavelength range (3600\,--\,10300\,\AA), AGN spectra are largely degenerate with that of the stellar populations \citep{CardosoAGNSED, LejaAGNSED}. To mitigate the influence of AGN contamination, we restrict our attention to low-luminosity, type 2 AGNs, which have a negligible effect on their host galaxy's optical SEDs. Although the MaNGA survey primarily contains type 2 AGNs, and the W18 selection criteria preferentially select low-luminosity AGNs, the AGN sample may nonetheless contain a few type 1 AGNs that are unsuitable for spatially resolved SED fitting. The spectra of galaxies hosting type 1 AGNs are characterized by broad emission lines. To identify low-redshift type 1 AGNs in SDSS DR7, \citet{OhType1} developed selection criteria that uses the FWHM of H$\alpha$ (FWHM H$\alpha\,{>}\,800$\,km\,s$^{-1}$) and the ratio of the H$\alpha$ Gaussian amplitude to the noise of H$\alpha$ (H$\alpha$ A/N\,$>$\,3). Analogously, to identify potential type 1 AGNs, we use spatially resolved H$\alpha$ Gaussian profile velocity, $\mathrm{Gvel}(\mathrm{H}\alpha)$, and the ratio of the H$\alpha$ Gaussian amplitude to H$\alpha$ noise in each spaxel, $\mathrm{A}(\mathrm{H}\alpha)/\mathrm{N}(\mathrm{H}\alpha)$. In analogy with the type 1 AGN criteria from \citet{OhType1} and the W18 selection criteria, we define two new spaxel parameters:

\begin{itemize}
\setlength\itemsep{0.15em}

\item \hspace{0.05em} $f_{\mathrm{A/N}}$: The $\mathrm{A}(\mathrm{H}\alpha)/\mathrm{N}(\mathrm{H}\alpha)$ spaxel fraction, i.e., the fraction of spaxels with an H$\alpha$ Gaussian amplitude to H$\alpha$ noise ratio greater than $10$\,\AA.

\item \hspace{0.05em} $f_{\mathrm{Gvel}}$: The $\mathrm{Gvel}(\mathrm{H}\alpha)$ spaxel fraction, i.e., the fraction of spaxels with an H$\alpha$ Gaussian profile velocity greater than $125$\,km\,s$^{-1}$.

\end{itemize}
The thresholds of $10$\,\AA\ and $125$\,km\,s$^{-1}$ were chosen based on the rightward tail of $\mathrm{Gvel}(\mathrm{H}\alpha)$ and $\mathrm{A}(\mathrm{H}\alpha)/\mathrm{N}(\mathrm{H}\alpha)$ for the 431 AGN galaxies. We adopt conservative cutoffs in $f_{\mathrm{A/N}}$ and $f_{\mathrm{Gvel}}$ to ensure no galaxies with significant AGN contamination sneak into the AGN sample, at the cost of a slight reduction in sample size. We use the following criteria to remove galaxies potentially hosting type 1 AGNs:

\begin{itemize}
\setlength\itemsep{0.15em}

\item $f_{\mathrm{A/N}}\,{>}\,40\,\%$ and $f_{\mathrm{Gvel}}\,{>}\,40\,\%$

\end{itemize}

Of the 302 face-on AGN candidates, 23 were excluded based on this cut, leaving a final AGN sample of 279 AGN galaxies. The 40\,\% cutoff fractions were chosen to be conservative --- W18 found a total of only 11 type 1 AGNs in their sample of 303 AGNs (by cross-referencing with \citealt{OhType1}). For a summary of the full AGN sample selection process, see Figure~\ref{fig:selectionmethod}.

Based on data from the SDSS NASA-Sloan Atlas (NSA), the mean stellar mass of the AGN sample is $\log($M$_{\ast}/$M$_\odot)\,{=}\,10.4$, with a mean redshift of $z\,{=}\,0.053$, as compared with a mean stellar mass of $\log($M$_{\ast}/$M$_\odot)\,{=}\,10.2$ and mean redshift of $z\,{=}\,0.046$ for all MaNGA DR16 galaxies (see Figure~\ref{fig:SMzdists}). Below, we select a sample of matched control galaxies against which we compare the AGN sample.

\subsection{Control Sample Selection}
\label{sec:controlsample}

\begin{figure*}
\centering
\includegraphics[width=0.9\textwidth]{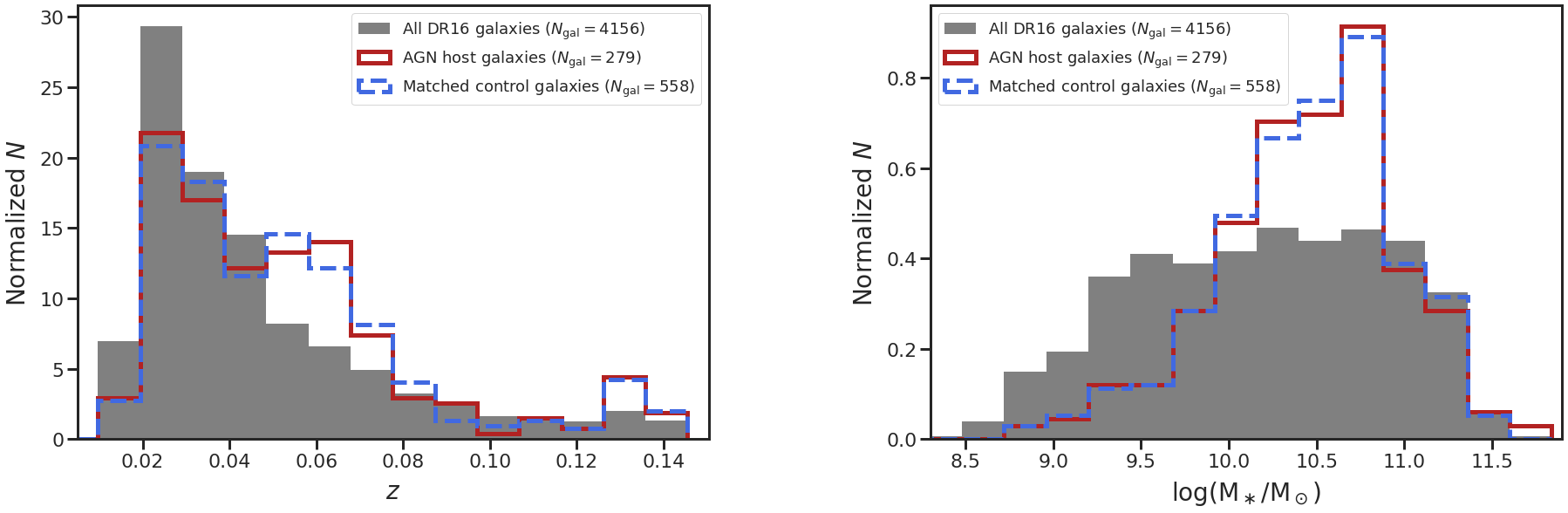}
\caption{Distribution of redshifts (left) and stellar masses (right) from the SDSS NSA for the AGN sample and the control sample, as compared with all MaNGA DR16 galaxies. The distribution of AGN galaxy redshifts is comparable to that of all MaNGA galaxies, whereas the stellar masses of the AGN galaxies are preferentially larger, peaking at $\log($M$_{\ast}/$M$_\odot)\,\sim\,10.5$. The control galaxy redshifts and stellar masses closely match those of the AGN galaxies (also match in $z$\,-\,$\log($M$_{\ast}/$M$_\odot)$ space).}
\label{fig:SMzdists}
\end{figure*}

To study the effect of AGN feedback on star formation, we adopt the approach of repeating our analyses on a sample of AGN-inactive matched control galaxies \citep[see also][]{CisternasAGNcontrol, EllisonAGNcontrol, KocevskiAGNcontrol, first62AGNI}. To compare against galaxies at a comparable evolutionary stage, but retain potentially informative differences in SFR, we select two stellar mass- and redshift-matched control galaxies. Before selecting the matched control galaxies, we apply the cuts performed on the AGN sample to the 3725 non-AGN galaxies (i.e., we remove edge-on and potential type 1 AGN galaxies). It is necessary to apply these cuts to the control sample to ensure they are not introducing an unintended bias to the AGN sample. Additionally, there may be type 1 AGNs missed by the W18 selection criteria present in the so-called non-AGN galaxies. With Galaxy Zoo data, we narrow the control sample to 2665 face-on galaxies. We remove a further 67 galaxies based on the spatially resolved type 1 AGN criteria (see Section~\ref{sec:AGNsample} for details), leaving a total of 2598 potential control galaxies (we compare morphologies and merger activity in Section~\ref{sec:morphs} below).

The full sample of potential control galaxies spans a range of stellar masses, redshifts, SFRs, etc. Because AGNs are known to reside preferentially in certain types of galaxies, directly comparing the properties of the AGN sample to the full sample of 2598 control galaxies is not necessarily a fair comparison. Instead, we study a control sample matched in stellar mass and redshift to the AGN galaxies. In detail, we first normalize the $\log($stellar masses$)$ and redshifts of all galaxies to a maximum of one. For each of the 279 AGN galaxies, we then select (without replacement) the two inactive galaxies which are closest in normalized stellar mass-redshift space and add them to the matched control sample. We weight the stellar mass by an arbitrary factor of $10$ in this selection to account for the lower variance in the distribution of normalized $\log($stellar masses$)$ as compared with the distribution of normalized redshifts. The resulting control sample of 558 galaxies closely matches the stellar mass and redshift distributions of the AGN host galaxies (see Figure~\ref{fig:SMzdists}), with a mean stellar mass of $\log($M$_{\ast}/$M$_\odot)\,{=}\,10.4$ and a mean redshift of $z\,{=}\,0.053$ (from NSA data). For a summary of the control sample selection process, see Figure~\ref{fig:selectionmethod}.

\subsection{SFH Reconstruction}
\label{sec:SFHfitting}

To study star formation trends over spatial and temporal scales, we aim to reconstruct nonparametric SFHs for MaNGA galaxies. Rather than re-perform SED fitting on MaNGA spectra, we turn to the extensively tested previous MaNGA SED fitting efforts. In particular, the PIPE3D analysis \citep{SanchezPipe3D1, SanchezPipe3D2} was used to derive stellar population data products for all MaNGA DR16 galaxies, and has provided the basis for a number of MaNGA-based studies. One relevant advantage of the PIPE3D analysis is that emission lines were not used to extract stellar population information, mitigating the possible influence of AGN contamination in our analysis. Previous works have used PIPE3D stellar population information to reconstruct simple piecewise-constant SFHs \citep[e.g.,][]{IbarraMedelMaNGAPIPE3D, SanchezMaNGAPIPE3D}. We instead reconstruct smooth, nonparametric SFHs with the help of PIPE3D data, following the dense basis methodology \citep{Iyerdb2}.

The PIPE3D pipeline was created based on the FIT3D fitting methodology, which fits spectra with stellar population models. FIT3D uses the simple stellar population library from \citet{FernandesSSPs}, which consists of 156 templates spanning 39 stellar ages (1\,Myr\,--\,14.1\,Gyr) and four metallicities (0.2\,Z/Z$_\odot$\,--\,1.5\,Z/Z$_\odot$). Before fitting spaxel spectra, the PIPE3D pipeline bins spaxels in the outskirts of galaxies, combining their spectra to achieve a desired S/N per \AA\ (specifically, 50, which gives an uncertainty of $\sim\,0.1$\,dex in the resulting stellar population properties). As a result, note that spaxels containing features like spiral arms may be smoothed over. Spaxels in the centers of MaNGA galaxies typically have an S/N per \AA\,${>}\,50$, and as a result, do not require any binning. In each spatial bin, spectra are co-added before a stellar population fit is performed. A multi-SSP linear fit is then carried out in each spatial bin using the library from \citet{FernandesSSPs}. Lastly, to resolve the stellar population properties in each MaNGA spaxel, a dezonification procedure is employed. This way, stellar population information is estimated in all spaxels, regardless of S/N. For more details about this process, see \citet{SanchezPipe3D2}.

For each spaxel in MaNGA galaxies, PIPE3D provides the fraction of luminosity produced by the SSPs across the range of stellar ages and metallicities (``luminosity fractions''). We convert the luminosity fractions to stellar mass fractions using the mass-to-light ratios provided for each SSP based on the \citet{SalpeterIMF} initial mass function. Stellar mass fractions indicate the times at which the spaxel has formed certain quantiles of its total stellar mass, placing constraints on the underlying SFH. We then reconstruct nonparametric SFHs with the dense basis approach.

The dense basis method uses Gaussian processes to create smooth SFHs independent of any functional form that connect stellar mass quantiles (from PIPE3D, in this case) in a physical way \citep{Iyerdb2}. That is, for stellar mass quantile times $\{t_x\}$, the reconstructed SFR(t) must satisfy
\begin{equation} 
\label{GPconstraints}
\int_{t=0}^{t=t_x}\,\mathrm{SFH}(t)\,f_{\mathrm{ret}}(t - t_x, Z)\,dt = M_{\ast}(t_x)
\end{equation}
, where $f_{\mathrm{ret}}(t - t_x, Z)$ is the metallicity-dependent fraction of the mass formed that is retained at the time of observation (typically between $0.6$\,--\,$1.0$; \citealt{ConroyFSPS}). We use Gaussian process regression to construct a smooth curve in stellar mass-lookback time space which passes through the constraint points (Eq.~\ref{GPconstraints}) provided by PIPE3D stellar population information. Differentiating the cumulative mass curve yields the SFH in the form of SFR($t$). For the Gaussian process regression, we adopt a physically motivated Matern32 kernel and never allow SFR to drop below zero in the smooth reconstruction (see \citealt{Iyerdb2} for more details). This Gaussian process-based description of SFHs minimizes the bias in reconstructed parameters (stellar mass, SFR, etc.) when compared to other parametric and nonparametric methods. Additionally, our reconstructed SFHs are robust to multiple episodes of star formation/quenching, which played an important role in our results. Stellar masses and SFRs derived from our reconstructed SFHs are consistent with the publicly released PIPE3D stellar masses and SFRs \citep{SanchezPipe3D2, Sanchez2022}.

\section{Host Galaxy Properties}
\label{sec:SFtrends}

\subsection{Galaxy Morphologies}
\label{sec:morphs}

Following \citet{WillettGZ2} and \citet{ComerfordAGNMaNGA}, we use Galaxy Zoo data to classify the morphology of the galaxies in a binary fashion: spiral or elliptical. Galaxies with a higher ``features or disk'' debiased vote fraction we term ``spiral'' and galaxies with a higher ``smooth'' debiased vote fraction we term ``elliptical''. Although this binary morphological classification is idealistic, in that not all galaxies can simply be classified as spiral/elliptical, it allows us to separate the AGN and control samples into two morphological bins for subsequent analyses. We also use Galaxy Zoo data to identify galaxies that are currently undergoing a merger.

Among the sample of AGN host galaxies, $81$\,\% are classified as spirals, $19$\,\% are classified as ellipticals, and $4.0$\,\% are classified as undergoing a merger. On the other hand, the matched control sample consists of $55$\,\% spiral galaxies and $45$\,\% elliptical galaxies, of which $3.4$\,\% are undergoing a merger (two control galaxies have inconclusive tied spiral and elliptical vote fractions). Our sample of optically-identified AGNs are preferentially found in spiral galaxies, with no obvious sign of merger-enhanced activity when compared with the matched control galaxies. This is consistent with the finding in previous studies that optical type 2 AGNs primarily reside in early-type spiral galaxies \citep[e.g.,][]{TorrecillaCALIFAAGN2, SanchezAGNSFMS}. To mitigate the impact of different morphologies in the AGN and control samples, we separate the AGN and control galaxies by morphology in many of the subsequent analyses (one could instead select control galaxies that are matched in morphology to the AGN galaxies, but it proves interesting to separate the samples by morphology). We explore star formation trends in the AGN and control samples below.

\subsection{Galaxy-Wide and Spatially Resolved SFRs}
\label{sec:SFRs}

\begin{figure}
\centering
\includegraphics[width=0.45\textwidth]{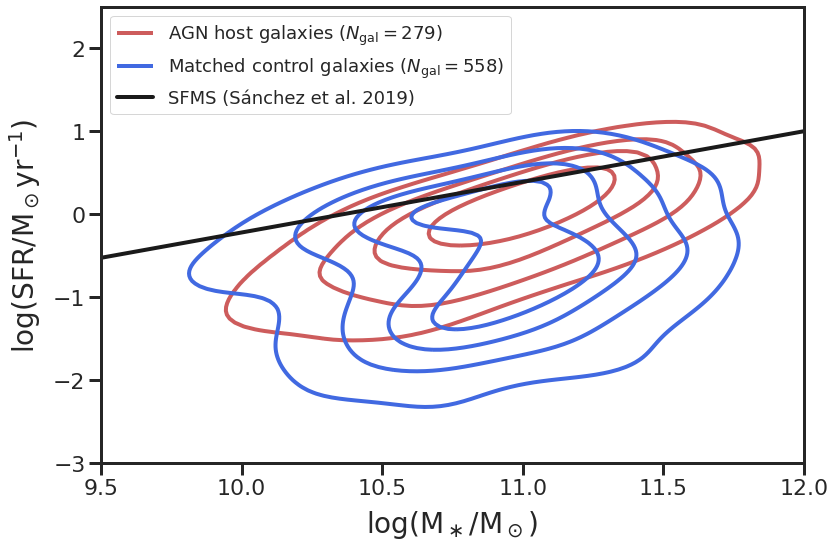}
\caption{Density contours in SFR-stellar mass space for the AGN host galaxies (red) and the control galaxies (blue), as compared with the SFMS from \citet{SanchezMaNGAPIPE3D}. The AGN galaxies predominantly lie on the SFMS/in the green valley region. By comparison, the control sample has a more broad distribution of SFRs, and a larger fraction of quiescent galaxies (i.e., galaxies 1.0\,dex below the SFMS; 23\,\% of the AGNs and 48\,\% of the control galaxies are quenched).}
\label{fig:SFMS}
\end{figure}

\begin{figure*}
\centering
\includegraphics[width=0.9\textwidth]{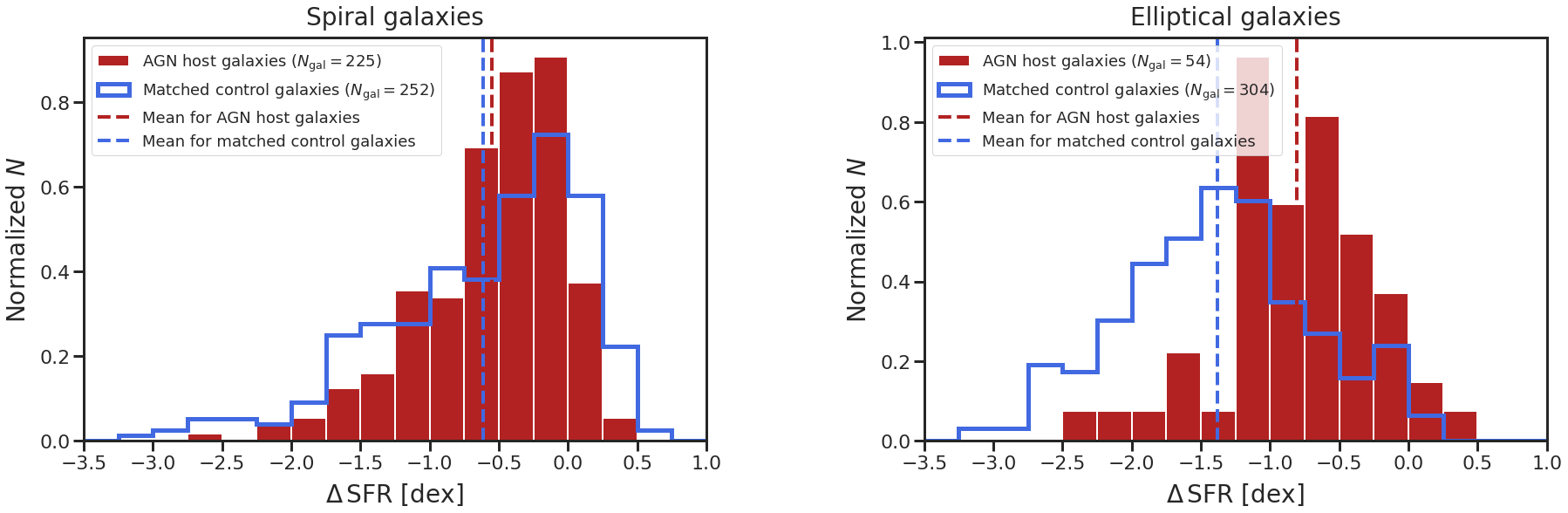}
\caption{Distribution of offsets from the SFMS for spiral AGN and control galaxies (left) and elliptical AGN and control galaxies (right). The AGN galaxies typically have larger SFRs than the matched control galaxies, with mean values highlighted as vertical lines. The SFRs of the AGN and control samples are more comparable among spiral galaxies (two-sample KS test 0.03 $p$-value) and differ substantially among elliptical galaxies (two-sample KS test $5.8\,\sigma$ significance). There is a population of quenched ($\Delta\,\mathrm{SFR}\, < 1.0$\,dex) elliptical galaxies that is seen in the control sample but not in the AGN sample.}
\label{fig:SFRoffsets}
\end{figure*}

Reconstructing spatially resolved SFHs allows us to analyze trends in both galaxy-wide and kiloparsec-scale star formation. Turning first to galaxy-wide trends, it is well established that the rate at which galaxies form stars is tied to their stellar mass. The observed linear relationship between $\log($SFR$)$ and $\log($stellar mass$)$ is termed the ``star formation main sequence'' \citep[SFMS e.g.,][]{BrinchmannSFMS, ElbazSFMS, SpeagleSFMS, Renzini&PengMS}. Despite much previous work, the location of AGN host galaxies relative to the SFMS remains somewhat unclear. Some studies have concluded that AGN host galaxies reside primarily in the green valley between star forming and quenched, whereas other works have found that AGN host galaxies lie distinctly on/above/below the SFMS \citep[e.g.,][]{SchawinskiAGNSFMS, RosarioAGNSFMS, YoungAGNSFMS, MullaneyAGNSFMS, StanleyAGNSFMS, SanchezAGNSFMS}. Additionally, some studies have found that the location of AGN galaxies on the SFR-stellar mass diagram depends on the type of AGN \citep[e.g.,][]{EllisonAGNSFMS, ComerfordAGNMaNGA}.

Throughout this paper, we use the mean SFR of the last $100$\,Myr of the spaxel/galaxy SFH (rather than, for instance, emission lines). This timescale is consistent with infrared tracers of SFR. Our sample of AGN galaxies and control galaxies is shown in comparison with the (PIPE3D) SFMS in Figure~\ref{fig:SFMS}. The AGN host galaxies largely reside on the SFMS, or slightly below, in the green valley region. By comparison, the control galaxies extend further above/below the SFMS than the AGN galaxies, and include a larger population of quenched galaxies (note that we use the terms ``quenched'' and ``quiescent'' interchangeably). Adopting a quenched threshold of 1.0\,dex below the SFMS, a total of 267 control galaxies are quiescent (48\,\%), whereas only 63 AGN galaxies are quiescent (23\,\%). This is not overly surprising; AGN activity is generally not anticipated in quiescent galaxies due to the lack of gas content required to fuel SMBH accretion and detect AGN activity. As such, we expect that the smaller proportion of quenched galaxies in our AGN sample is caused by the relative rarity of AGN activity in the quiescent galaxy population, rather than, for instance, positive AGN feedback.

It is worth noting that the galaxy-wide stellar masses shown in Figure~\ref{fig:SFMS}, measured with spatially resolved SED fitting, do not exactly agree with the NSA stellar masses derived from full-galaxy SDSS SEDs, which were used to match stellar masses during sample selection (Figure~\ref{fig:SMzdists}). We find that the NSA stellar mass are underestimated by ${\sim}\,0.3$\,dex, with a fair amount of scatter. As a result, the reconstructed stellar masses of the AGN host galaxies and the matched control galaxies are no longer precisely one-to-one, but they remain largely consistent (with a two-sample KS test $p$-value of $0.22$). The discrepancy between the stellar masses derived from unresolved SEDs and spatially resolved SEDs has been previously studied \citep{ZibettiSMSED, SorbaSMSED1, SorbaSMSED2}, and in this case, different choices of the initial mass function between PIPE3D and the NSA catalog likely also contributed.

\begin{figure*}
\centering
\includegraphics[width=0.9\textwidth]{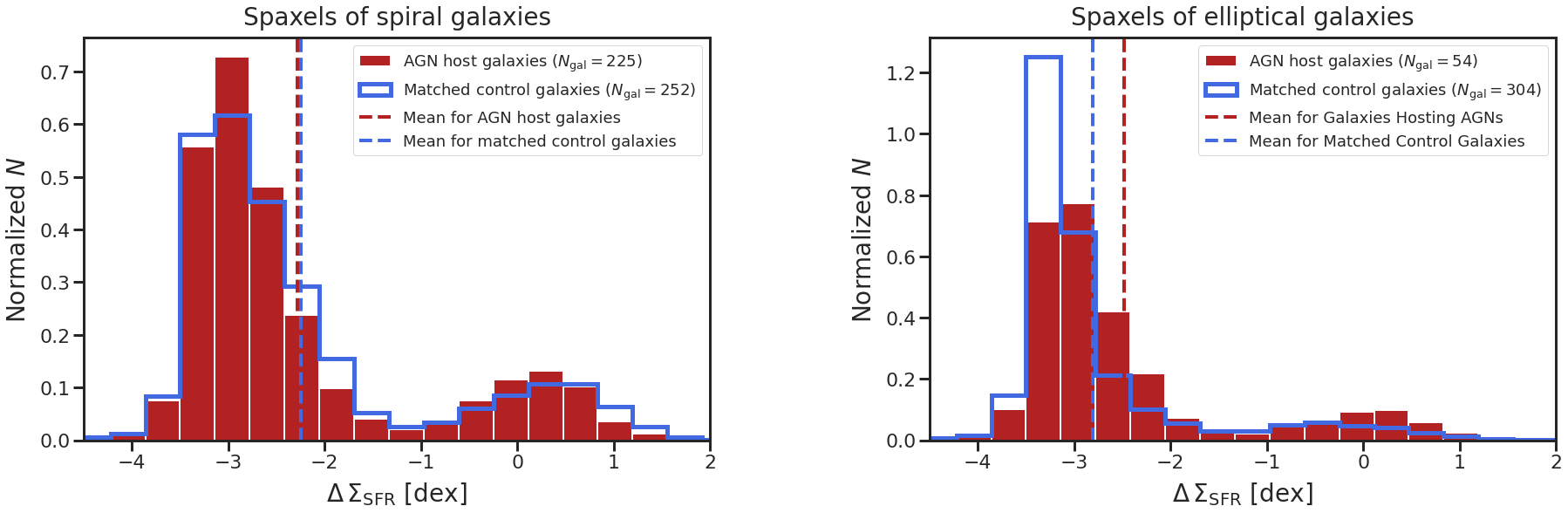}
\caption{Distribution of spaxel offsets from the local SFMS for spiral AGN and control galaxies (left) and elliptical AGN and control galaxies (right). Similarly to Figure~\ref{fig:SFRoffsets}, the distributions of spaxel SFR densities in the spiral AGN and spiral control galaxies match more closely than the distributions of SFR densities in the elliptical AGN and elliptical control galaxies. This plot also illustrates that typical galaxies, both quiescent ($\Delta\,\mathrm{SFR}\, < 1.0$\,dex) and nonquiescent, contain many quenched spaxels.}
\label{fig:spaxelSFRoffsets}
\end{figure*}

\begin{figure*}
\centering
\includegraphics[width=\textwidth]{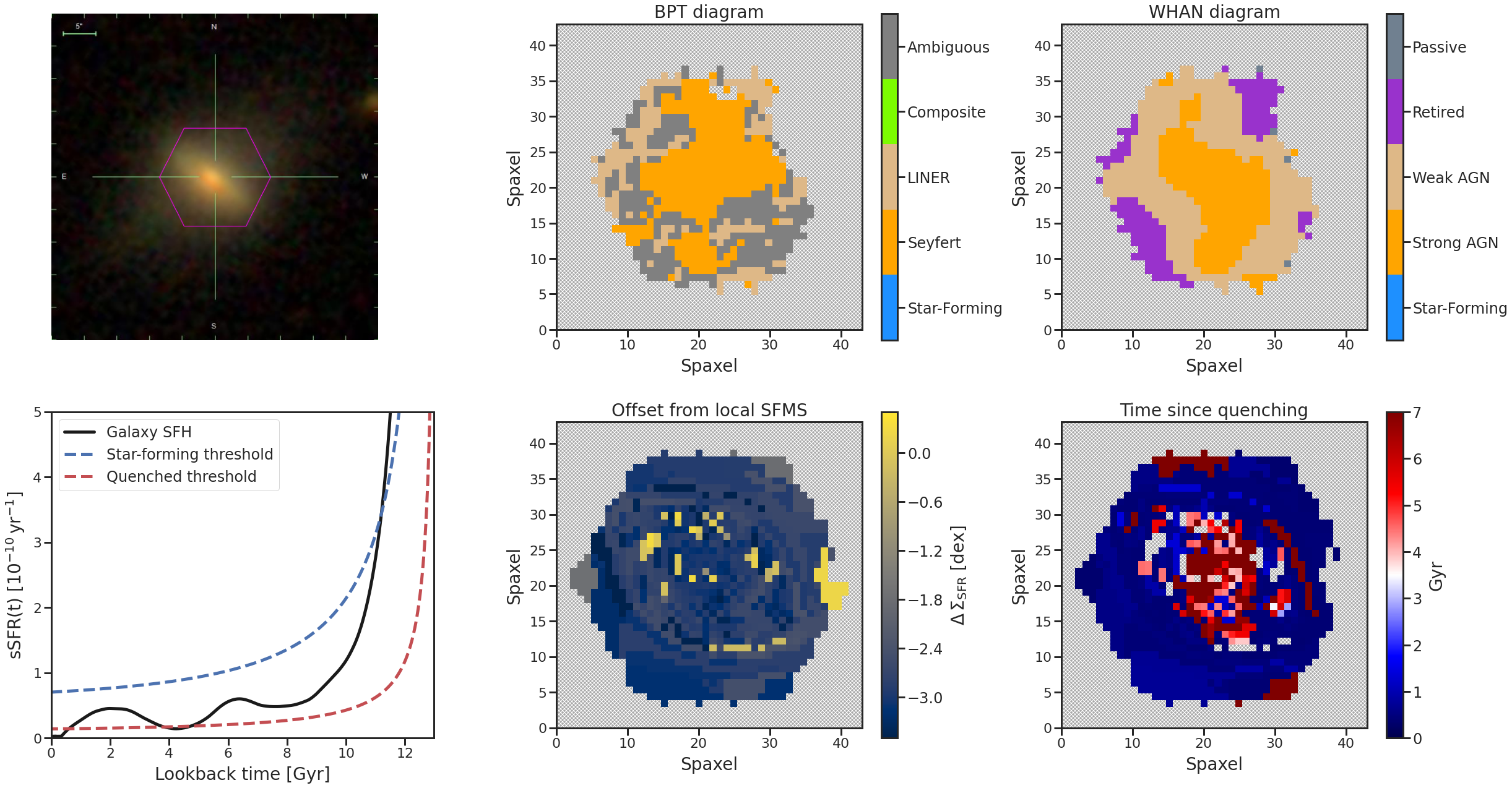}
\caption{Plots for MaNGA AGN galaxy 1-137883 that illustrate trends in the AGN sample. The top row shows the optical image with MaNGA hexagon coverage overlay, and the spatially resolved BPT and WHAN diagrams. The bottom row shows the galaxy's sSFR as a function of lookback time with highlighted star forming and quenched thresholds ($\mathrm{sSFR}\,{=}\,1.0/t_{H}$ and $\mathrm{sSFR}\,{=}\,0.2/t_{H}$, respectively), along with spatially resolved diagrams of ${\Delta}\,{\Sigma}_{\mathrm{SFR}}$ and time since quenching. The time since quenching plot shows the quenching time of spaxels that are quenched at the time of observation (non-quenched spaxels are shown in gray). Star formation in the central regions of the galaxy quenched early-on (${\sim}\,6$\,Gyr ago). In this case, galaxy-wide star formation also quenched, instead taking place over the last ${\sim}\,1$\,Gyr.}
\label{fig:example}
\end{figure*}

Upon further investigation, the SFRs of the AGN and control galaxies, and thereby their positions relative to the SFMS, strongly depend on galaxy morphology. Breaking up the AGN and control galaxies by morphology reveals, unsurprisingly, that elliptical galaxies generally reside lower with respect to the SFMS than spiral galaxies. Figure ~\ref{fig:SFRoffsets} shows the distribution of offsets from the SFMS, defined as
\begin{equation}
\label{SFRoffset}
\Delta\,\mathrm{SFR} = \log(\mathrm{SFR}_{\mathrm{gal}})\,{-}\,\log(\mathrm{SFR}_{\mathrm{SFMS}})
\end{equation}
. Overall, the AGN galaxies have slightly larger SFRs than the control galaxies. The differences in SFRs between the AGN and control galaxies are more significant among ellipticals than they are among spirals; elliptical control galaxies typically have lower $\Delta\,\mathrm{SFR}$ than elliptical AGN galaxies (the two-sample KS test gives a $5.8\,\sigma$ significance). On the other hand, the distribution of $\Delta\,\mathrm{SFR}$ for spiral AGN galaxies and spiral control galaxies are more comparable, with a marginally consistent $p$-value of 0.03 from the two-sample KS test and a similar mean. In other words, AGNs hosted in spiral galaxies have fairly typical spiral galaxy SFRs, whereas AGNs hosted in elliptical galaxies have preferentially larger SFRs than typical ellipticals. Again, we expect that this is caused by the necessity of gas content to fuel AGN activity; for ellipticals to host AGNs they must contain a reservoir of gas, possibly due to recent mergers, which could also fuel star formation (indeed, trends in $\Delta\,\mathrm{SFR}$ are mirrored in the distributions of gas masses, as provided by PIPE3D). Aside from the smaller fraction of spiral AGN galaxies above the SFMS ($\Delta\,\mathrm{SFR}\,{>}\,0$) when compared with the spiral control galaxies, we find no glaring signatures of AGN feedback in the galaxy-wide SFRs.

Large IFU galaxy surveys, such as CALIFA and MaNGA, have revealed that spatially resolved spaxels also follow a tight relationship between $\log($SFR$)$ and $\log($stellar mass$)$, termed the spatially resolved SFMS or the local SFMS \citep[e.g.,][]{SanchezCALIFAMM, CanoresolvedSFMS, HsiehresolvedSFMS, Bluckquenching, SanchezresolvedSFMS, BakerresolvedSFMS}. The local SFMS describes a linear relationship between $\log($SFR density$)$ and $\log($stellar mass density$)$ to reflect the varying physical size of individual spaxels. Further investigating star formation trends, Figure~\ref{fig:spaxelSFRoffsets} shows the distribution of offsets from the local SFMS in individual spaxels, defined analogously to $\Delta\,\mathrm{SFR}$:
\begin{equation} 
\label{localSFRoffset}
\Delta\,\Sigma_{\mathrm{SFR}} = \log(\Sigma_{\mathrm{SFR},\,\mathrm{spaxel}})\,{-}\,\log(\Sigma_{\mathrm{SFR},\,\mathrm{SFMS}})
\end{equation}
. Adopting the local SFMS from \citet{Bluckquenching}, we see a bimodal distribution of star-forming and quenched spaxels, with many quenched spaxels in both spirals and ellipticals. The absence of spaxels in the green valley region is due to the relatively rapid spaxel quenching times ($<\,1$\,Gyr) in our reconstructed SFHs. As we found for $\Delta\,\mathrm{SFR}$, the distributions of $\Delta\,\Sigma_{\mathrm{SFR}}$ are comparable for spiral AGNs and spiral control galaxies, whereas the distributions of $\Delta\,\Sigma_{\mathrm{SFR}}$ for elliptical AGNs and elliptical control galaxies differ more substantially.

Based on galaxy-wide SFRs, as well as the distribution of spaxel SFR densities, we see no obvious signatures of AGN feedback in our sample of AGN galaxies. Instead, we find that the AGN sample has a smaller spread of galaxy SFRs and a missing population of quenched (primarily elliptical) galaxies. However, AGN feedback may still have a substantial effect on star formation in the central regions of the AGN galaxies (see, e.g., Figure~\ref{fig:example}). We explore this possibility below.

\subsection{Radial Star Formation Profiles}
\label{sec:radialSFRs}

Observations of gas content in low-redshift AGN galaxies suggest that AGN feedback may primarily take place in the central (kiloparsec-scale) regions of galaxies \citep{SanchezAGNSFMS, EllisonAGNgas}. Here, we test whether star formation trends in the central regions of AGN galaxies differ from those of matched control galaxies. Previous studies have analyzed the radial star formation trends of large galaxy samples in MaNGA, revealing that SFR is enhanced and quenched from the inside out \citep[e.g.,][]{BelfioreradialSFR, EllisonradialSFR, SanchezAGNSFMS, Bluckquenching}. Following the previous works, we display radial SFR profiles in terms of the offset from the local SFMS (${\Delta}\,{\Sigma}_{\mathrm{SFR}}$), thereby accounting for the SFR expected due to the stellar mass contained in each spaxel. We bin spaxels in elliptical apertures and calculate the mean ${\Delta}\,{\Sigma}_{\mathrm{SFR}}$ across all galaxies in the AGN and control samples. Note that spaxels with an SFR density of exactly $0$ are necessarily excluded when calculating the mean ${\Delta}\,{\Sigma}_{\mathrm{SFR}}$ (but ${\sim}\,0$\,\% of spaxels have an SFR density of exactly $0$). To emphasize trends in the central regions (and where these trends disappear), we plot the \textit{cumulative} mean ${\Delta}\,{\Sigma}_{\mathrm{SFR}}$ of spaxels within distance $R$.

\begin{figure}
\centering
\includegraphics[width=0.45\textwidth]{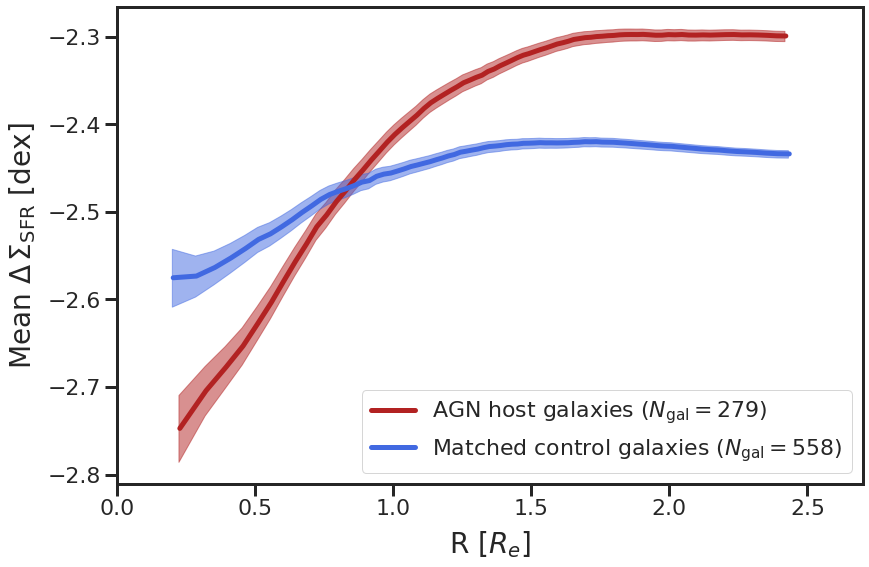}
\caption{Mean offset from the local SFMS for spaxels within $R$ (i.e., $R\,{=}\,0.5\,R_e$ shows the mean offset of spaxels across all galaxies located within $R\,{<}\,0.5\,R_e$). The shaded regions show the 3\,$\sigma$ error on the mean. Despite greater overall SFR densities, the AGN galaxies have lower central SFR densities when compared with the control galaxies.}
\label{fig:radialSFRs}
\end{figure}

\begin{figure*}
\centering
\includegraphics[width=0.9\textwidth]{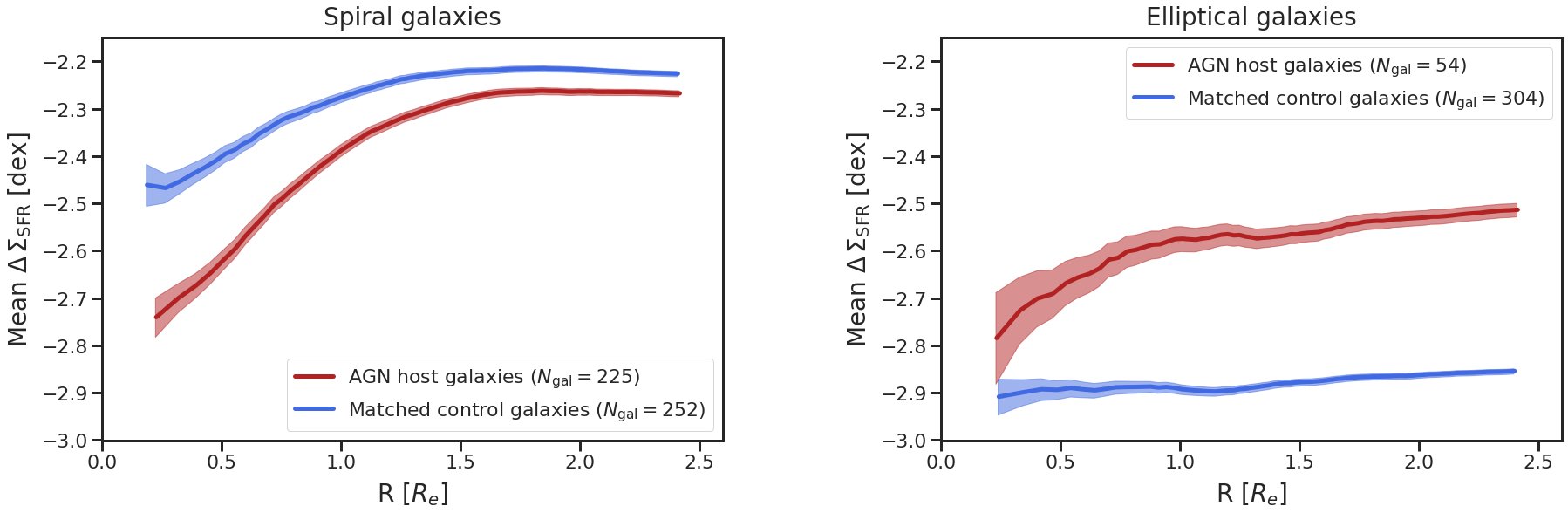}
\caption{Mean offset from the local SFMS for spaxels within $R$ in spiral AGN and control galaxies (left) and elliptical AGN and control galaxies (right). Central SFR densities are significantly lower in the spiral AGN galaxies when compared with the spiral control galaxies. Among elliptical galaxies, the AGN galaxies have larger overall SFR densities but still display a more inside out profile. Notice that the right-most mean offsets agree with with the mean offsets shown in Figure~\ref{fig:spaxelSFRoffsets} (as they must).}
\label{fig:radialSFRsmorph}
\end{figure*}

\begin{figure*}
\centering
\includegraphics[width=0.9\textwidth]{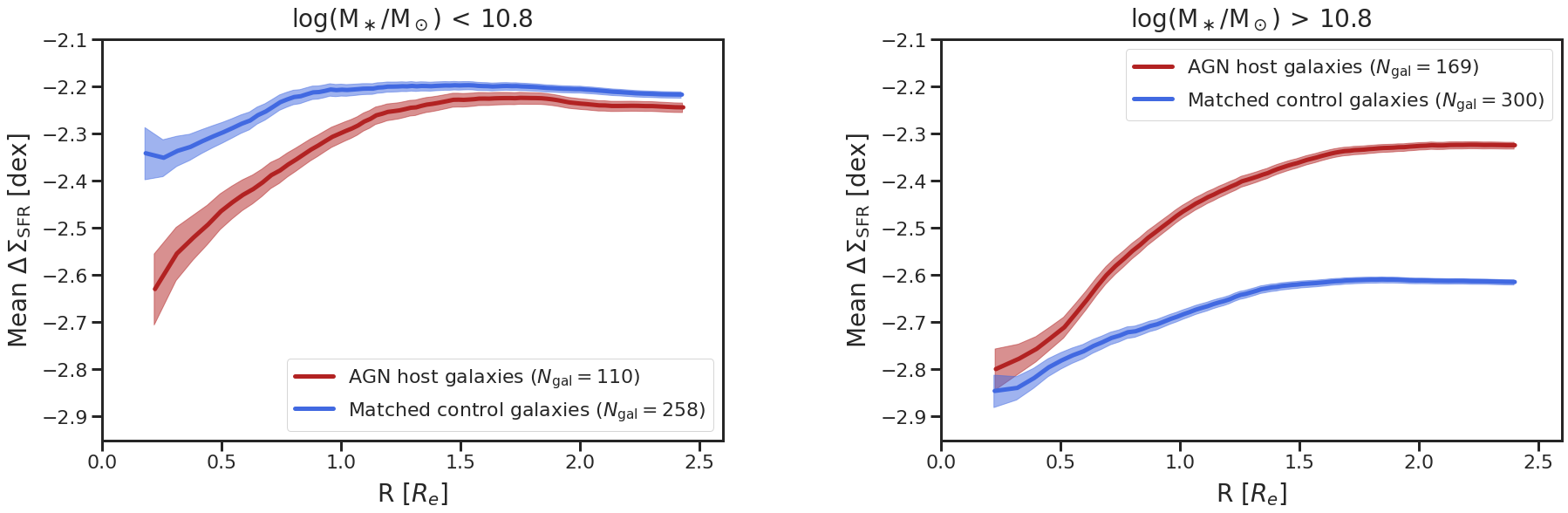}
\caption{Mean offset from the local SFMS for spaxels within $R$ in AGN galaxies and control galaxies with $\log($M$_{\ast}/$M$_\odot)\,{<}\,10.8$ (left) and $\log($M$_{\ast}/$M$_\odot)\,{>}\,10.8$ (right). Both lower-stellar and higher-stellar mass AGN galaxies have centrally suppressed SFR densities. Higher-stellar mass control galaxies display a qualitatively comparable central suppression, however, lower-stellar mass control galaxies have a more flat profile.}
\label{fig:radialSFRsSM}
\end{figure*}

Figure~\ref{fig:radialSFRs} shows the radial ${\Delta}\,{\Sigma}_{\mathrm{SFR}}$ profiles for the AGN galaxies and the control galaxies. When compared with the control galaxies, we find that the AGN galaxies have a larger overall mean ${\Delta}\,{\Sigma}_{\mathrm{SFR}}$, consistent with trends in galaxy-wide SFRs. However, the AGN galaxies have centrally suppressed ${\Delta}\,{\Sigma}_{\mathrm{SFR}}$ within $R\,{\lessapprox}\,0.5\,R_e$. This provides observational evidence of AGN feedback suppressing star formation in the central regions of low-redshift galaxies. The difference between the mean central ${\Delta}\,{\Sigma}_{\mathrm{SFR}}$ of the AGN and control samples is relatively small (${\sim}$\,0.2\,dex at the smallest $R$); however, this is a statistically significant effect that emerges from tens-to-hundreds of thousands of spaxels (depending on $R$), resulting in small uncertainties on the mean. This indicates that AGN galaxies have suppressed central star formation when compared with the control galaxies, providing evidence of AGN feedback hindering star formation. Breaking up the radial ${\Delta}\,{\Sigma}_{\mathrm{SFR}}$ profiles by galaxy morphology helps to clarify the trends.

Figure~\ref{fig:radialSFRsmorph} shows the radial ${\Delta}\,{\Sigma}_{\mathrm{SFR}}$ profiles for the AGN and control galaxies, separated into spirals and ellipticals. Among spirals, the AGN galaxies have a centrally suppressed ${\Delta}\,{\Sigma}_{\mathrm{SFR}}$ profile, lying ${\sim}\,0.3$\,dex below the control galaxies (factor of 2) at the smallest $R$. Again, this arises due to differences in the SFR densities of tens-to-hundreds of thousands of spaxels (Figure~\ref{fig:spaxelSFRoffsets} shows the corresponding SFR density distributions for all spaxels). The difference between the spiral AGN and control ${\Delta}\,{\Sigma}_{\mathrm{SFR}}$ profiles narrows as $R$ approaches 2.5. Among elliptical galaxies, the AGN galaxies instead have larger overall ${\Delta}\,{\Sigma}_{\mathrm{SFR}}$. However, the radial ${\Delta}\,{\Sigma}_{\mathrm{SFR}}$ profile of the AGN galaxies display central SFR suppression that is not found for the control galaxies (notice that the right-most mean ${\Delta}\,{\Sigma}_{\mathrm{SFR}}$ in these profiles agrees with the mean ${\Delta}\,{\Sigma}_{\mathrm{SFR}}$ shown in Figure~\ref{fig:spaxelSFRoffsets}). This explains the larger SFR densities in the outskirts of the AGN galaxies when compared with the control galaxies in Figure~\ref{fig:radialSFRs}. Furthermore, this demonstrates that suppressed central SFRs are responsible for the lack of spiral AGN galaxies above the SFMS, placing them instead in the green valley (see Figure~\ref{fig:SFRoffsets}).

\begin{figure*}
\centering
\includegraphics[width=0.9\textwidth]{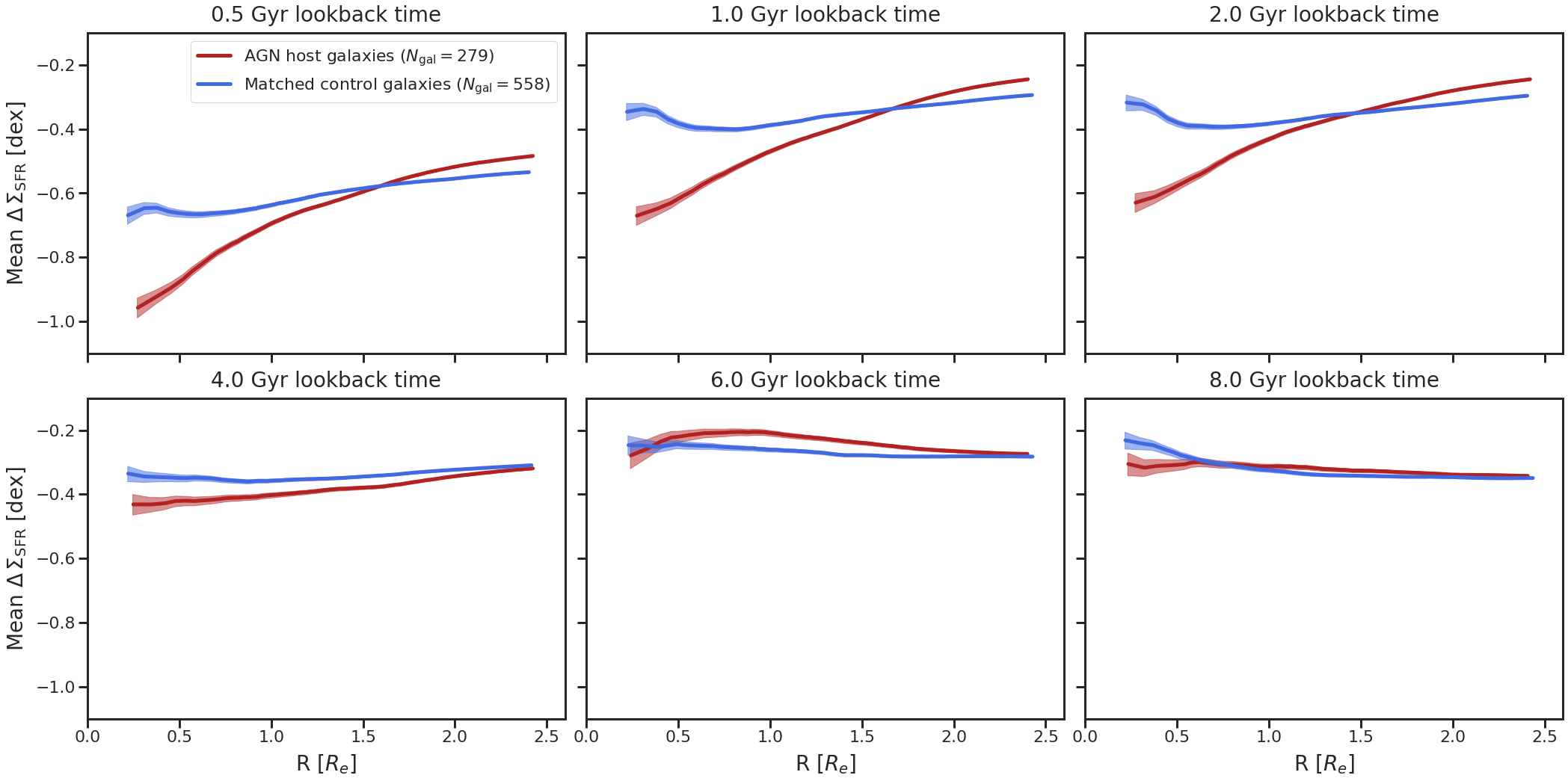}
\caption{Mean offset from the local SFMS for spaxels within $R$ for the AGN galaxies and control galaxies at lookback times ranging from $0.5$\,--\,$8.0$\,Gyr. At a lookback time of $\sim\,6$\,Gyr, suppression of the central SFR begins in the AGN galaxies, increasing over the subsequent Gyrs. Beyond $\sim\,6$\,Gyr, the radial profiles of the AGN and control galaxies are largely consistent. The consistency between the overall mean offsets of the AGN and control galaxies indicates that the AGN galaxies are not preferentially in the process of galaxy-wide quenching (trends are similar when excluding ellipticals).}
\label{fig:radialSFRslookback}
\end{figure*}

Due to the tight relationship between SMBH mass and galaxy stellar mass, AGN feedback is generally believed to be most significant in high-stellar mass galaxies \citep{SchawinskiAGNhighSM, WangAGNhighSM, ReinesBHmasses, BongiornoAGNhighSM, ForsterSchreiberoutflow}. As a result, many cosmological simulations implement AGN feedback primarily for high-stellar mass galaxies \citep[e.g.,][]{BoothAGNsim, VogelsbergerAGNsim, WeinbergerAGNsim}. When we separate the radial ${\Delta}\,{\Sigma}_{\mathrm{SFR}}$ profiles by stellar mass, we find that trends also depend on stellar mass (Figure~\ref{fig:radialSFRsSM}). The ${\Delta}\,{\Sigma}_{\mathrm{SFR}}$ profiles for both higher-stellar mass AGN galaxies and higher-stellar control galaxies are inside out. Among lower-stellar mass galaxies, the AGN ${\Delta}\,{\Sigma}_{\mathrm{SFR}}$ profile is centrally suppressed, whereas the control profile is more flat. This suggests that AGN feedback also occurs in lower-stellar mass galaxies. Additionally, as found above for spirals, this establishes that suppressed central SFRs are responsible for the position of the lower-stellar mass AGNs in the green valley (see Figure~\ref{fig:SFRoffsets}). Unfortunately, we are limited by the small number of low-stellar mass AGNs in our sample, but the trends of the few lowest-stellar mass galaxies ($\log($M$_{\ast}/$M$_\odot)\,{<}\,10.0$) are comparable to the left panel of Figure~\ref{fig:radialSFRsSM}.

\subsection{Evolution of Radial Star Formation Profiles}
\label{sec:evolradialSFRs}

Radial ${\Delta}\,{\Sigma}_{\mathrm{SFR}}$ profiles, although useful for gaining insight into radial star formation trends, provide no timescale information. In the analysis above, the timescale over which central SFR suppression arises in the AGN galaxies is not clear, which raises the question: is central quenching occurring on Myr-timescales (suggesting recent outflows) or Gyr-timescales (suggesting longer-term, integrated AGN feedback)? With reconstructed SFHs, we are able to address this question. We create analogous radial SFR plots to Figure~\ref{fig:radialSFRs} at lookback times spanning the last 8\,Gyr to study the evolution of ${\Delta}\,{\Sigma}_{\mathrm{SFR}}$ profiles over cosmic time.

One complication to creating ${\Delta}\,{\Sigma}_{\mathrm{SFR}}$ profiles at specific lookback times is that the local SFMS also changes over time. Despite several studies on the evolution of the global SFMS over cosmic time \citep[e.g.,][]{SpeagleSFMS, PueblaSFMS, SanchezMaNGAPIPE3D, PopessoSFMS}, the evolution of the local SFMS has not been studied quantitatively. Lacking a quantitatively determined local SFMS, we instead adopt a local SFMS of $\mathrm{SFR}(t)\,{=}\,M_{\ast}/t_{H}(t)$, where $t_{H}(t)$ is the age of the Universe at lookback time $t$, in analogy with the simplified, time-parameterized global SFMS adopted in previous works \citep[e.g.,][]{PacificiSFHs, CarnallSFHs, MonteroSIMBA, TacchellaSFHs}. At the time of observation (i.e., $t$\,{=}\,$0$), this choice closely agrees with the local SFMS from \citet{Bluckquenching}. Because we are primarily interested in comparing the AGN galaxies to the control galaxies, the accuracy of the adopted local SFMS is not crucially important.

Figure~\ref{fig:radialSFRslookback} shows snapshots of the AGN and control ${\Delta}\,{\Sigma}_{\mathrm{SFR}}$ profiles at lookback times spanning $0.5$\,--\,$8.0$\,Gyr. The central SFR suppression in the AGN galaxies began ${\sim}\,6$\,Gyr ago, increasing over the next few gigayears and resulting in significant central suppression that is already present by ${\sim}\,2$\,Gyr ago. Beyond ${\sim}\,6$\,Gyr ago, the AGN and control galaxies have a comparably flat ${\Delta}\,{\Sigma}_{\mathrm{SFR}}$ profile. Trends are similar if we exclude elliptical galaxies from this plot (over concerns of stars that were formed ex-situ).

The overall consistency between the radial AGN and control ${\Delta}\,{\Sigma}_{\mathrm{SFR}}$ profiles over the last ${\sim}\,2$\,Gyr indicates that the AGN galaxies are not preferentially in the process of galaxy-wide quenching when compared with the matched control galaxies. If this were the case, we would expect the overall mean spaxel SFRs of the AGN galaxies to decrease relative to that of the control galaxies over time, whereas they remain largely consistent (barring the central suppression). Instead, we find that the AGN galaxies have often undergone quenching in their central regions several gigayears ago that did not lead to galaxy-wide quenching.

Due to the bimodal nature of the ${\Delta}\,{\Sigma}_{\mathrm{SFR}}$ distributions (Figure~\ref{fig:spaxelSFRoffsets}), mean ${\Delta}\,{\Sigma}_{\mathrm{SFR}}$ profiles always reside below zero. On a related note, notice that the y-axes of Figure~\ref{fig:radialSFRs} and Figure~\ref{fig:radialSFRslookback} differ. This is a numerical artifact of the reconstructed SFHs: at $t\,{=}\,0$, very few quenched spaxels have an SFR density of exactly $0$ (see Figure~\ref{fig:spaxelSFRoffsets}), but at larger lookback times, a more substantial fraction of the quenched spaxels have an SFR density of exactly $0$ (this is caused by the limited time resolution of the SFHs, see Section~\ref{sec:caveats}). These quenched spaxels are necessarily excluded from the ${\Delta}\,{\Sigma}_{\mathrm{SFR}}$ distribution, thereby artificially increasing mean ${\Delta}\,{\Sigma}_{\mathrm{SFR}}$. At all lookback times, the AGN and control galaxies have a comparable fraction of quenched spaxels with an SFR density of exactly $0$, making Figure~\ref{fig:radialSFRslookback} a fair comparison.

\section{Star Formation Quenching}
\label{sec:qbehavior}

In Section~\ref{sec:SFtrends}, we found that the AGN galaxies have suppressed central SFRs, which arose over Gyr-timescales (beginning ${\sim}\,6$\,Gyr ago). To further quantify the central quenching timescale in the AGN galaxies, we introduce a radial metric that is complementary to mean ${\Delta}\,{\Sigma}_{\mathrm{SFR}}$: the mean time since quenching. This is a useful metric because even nonquiescent galaxies have many quiescent spaxels (Figure~\ref{fig:spaxelSFRoffsets}), which can be used to study quenching trends. Like radial profiles of ${\Delta}\,{\Sigma}_{\mathrm{SFR}}$, radial profiles of the time since quenching provide insights into qualitative quenching trends (inside out vs. outside in), with the additional advantage of directly quantifying the timescale over which quenching occurs.

\subsection{Radial Quenching Profiles}
\label{sec:radialqs}

In analogy with the specific star formation rate (sSFR)-based galaxy quenching definitions used in other works \citep[e.g.,][]{PacificiSFHs, MonteroSIMBA, CarnallSFHs, TacchellaSFHs}, we adopt a quenched spaxel threshold of $\mathrm{sSFR}(t)\,{<}\,0.2/t_{H}(t)$, which lies well below the local SFMS adopted in Section~\ref{sec:evolradialSFRs} ($\mathrm{sSFR}(t)\,{=}\,1.0/t_{H}(t)$). The time at which a spaxel quenches is defined as the time at which its sSFR drops below $0.2/t_{H}(t)$ (our conclusions are insensitive to the exact choice of quenching threshold). For straightforward interpretation, we restrict our analysis of radial quenching profiles to spaxels that are quenched at the time of observation.

\begin{figure}
\centering
\includegraphics[width=0.45\textwidth]{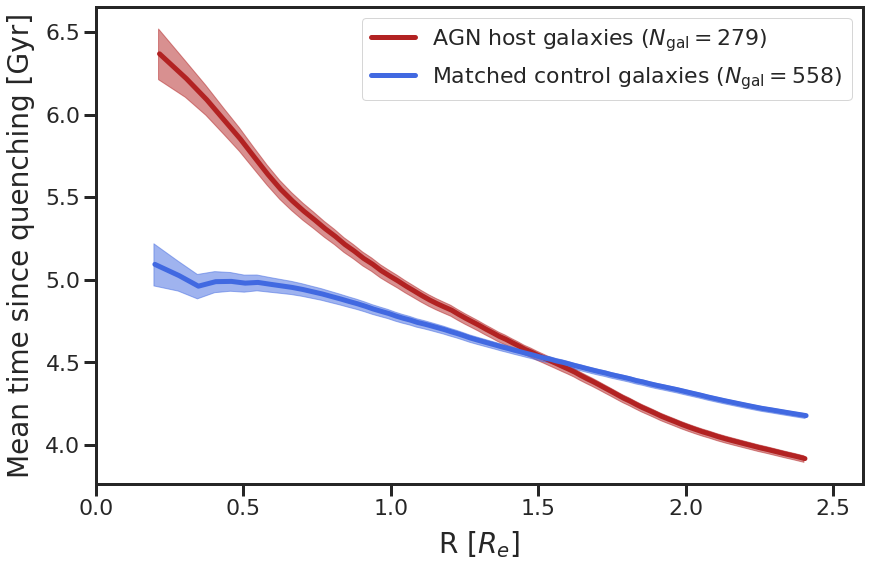}
\caption{Mean time since quenching of spaxels within $R$ (i.e., $R\,{=}\,0.5\,R_e$ shows the mean time since quenching of spaxels across all galaxies located within $R\,{<}\,0.5\,R_e$). The shaded regions show the 3\,$\sigma$ error on the mean. The AGN host galaxies display a more inside out radial quenching profile than the control galaxies, with quenching beginning in the center $6.5$\,Gyr ago. Notice the qualitative agreement with Figure~\ref{fig:radialSFRs}.}
\label{fig:radialtimeqs}
\end{figure}

Figure~\ref{fig:radialtimeqs} shows the radial time since quenching profile for the AGN and control galaxies. The AGN galaxies have a more steep inside out quenching profile than the control galaxies, with quenching beginning in the center $6.5$\,Gyr ago and taking place in the outskirts $2.5$\,Gyr later. Although quenching does occur in the outskirts of the AGN galaxies, note that this may not be driven by AGN feedback (as suggested by the results in Section~\ref{sec:SFtrends}), so this timescale should be interpreted accordingly. In the control galaxies, quenching instead began in the central regions $5$\,Gyr ago, spreading to the outskirts in $0.5$\,Gyr. Notice the qualitative agreement between the radial quenching profiles (Figure~\ref{fig:radialtimeqs}) and the radial ${\Delta}\,{\Sigma}_{\mathrm{SFR}}$ profiles (Figure~\ref{fig:radialSFRs}), which holds for all plots in this section. The agreement reinforces the usefulness of ${\Delta}\,{\Sigma}_{\mathrm{SFR}}$ profiles in gaining insight into qualitative Gyr-timescale radial quenching trends, as well as the quantitative advantage of radial quenching profiles.

\begin{figure*}
\centering
\includegraphics[width=0.9\textwidth]{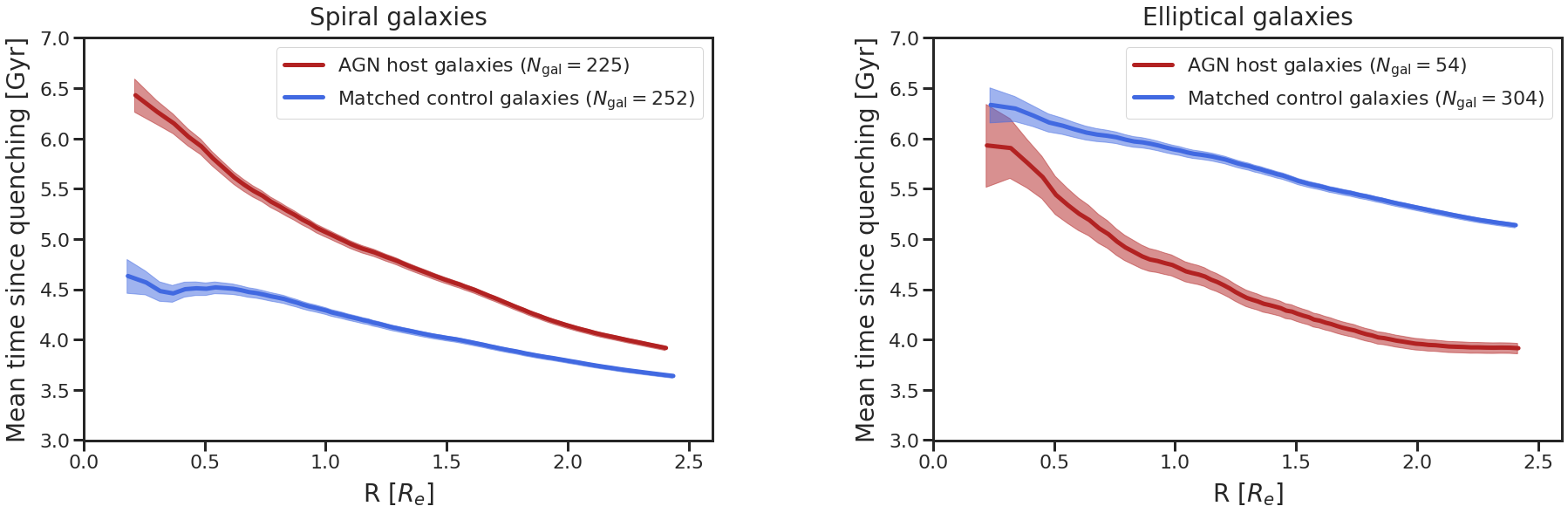}
\caption{Mean time since quenching of spaxels within $R$ in spiral AGN and control galaxies (left) and elliptical AGN and control galaxies (right). Spiral AGN galaxies display a significantly more inside out quenching profile than spiral control galaxies. Similarly, the inside out quenching profile of the elliptical AGN galaxies differs substantially from that of the elliptical control galaxies. Notice the qualitative agreement with Figure~\ref{fig:radialSFRsmorph}.}
\label{fig:radialtimeqmorph}
\end{figure*}

\begin{figure*}
\centering
\includegraphics[width=0.9\textwidth]{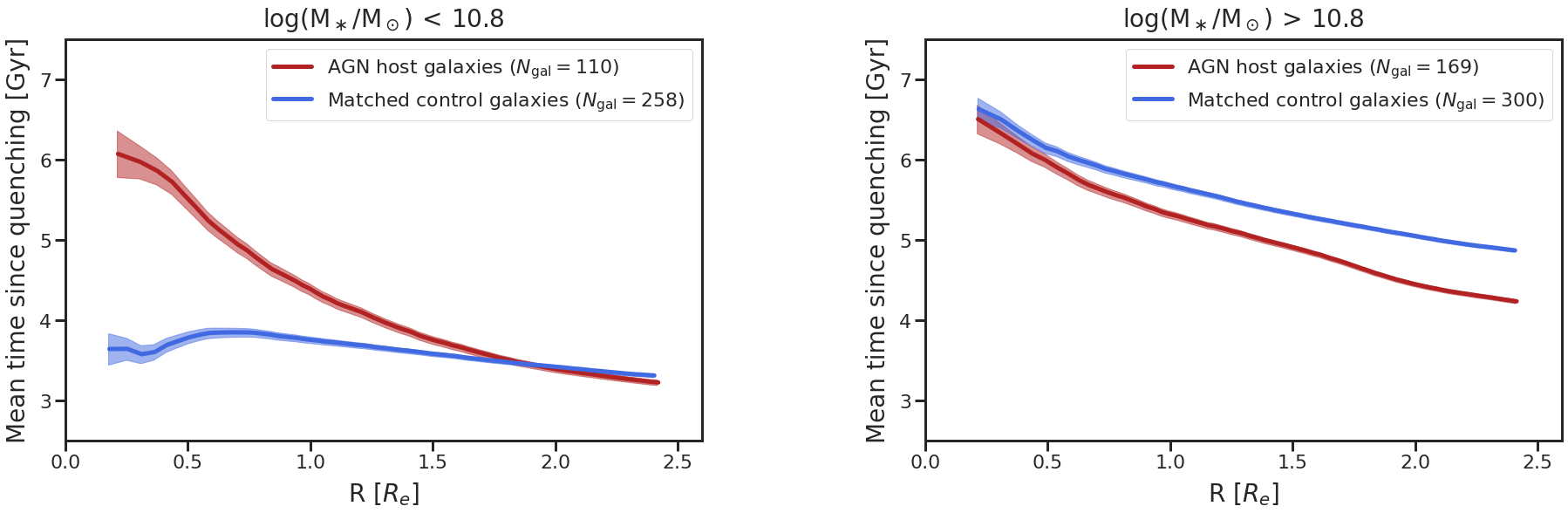}
\caption{Mean time since quenching of spaxels within $R$ in AGN and control galaxies with $\log($M$_{\ast}/$M$_\odot)\,{<}\,10.8$ (left) and $\log($M$_{\ast}/$M$_\odot)\,{>}\,10.8$ (right). Both higher-stellar mass and lower-stellar mass AGN galaxies have an inside out quenching profile, with reasonable quantitative agreement. Higher-stellar mass control galaxies display a similar inside out quenching profile, but the lower-stellar mass control galaxies instead have a flat quenching profile. Notice the qualitative agreement with Figure~\ref{fig:radialSFRsSM}.}
\label{fig:radialtimeqSM}
\end{figure*}

As in Section~\ref{sec:SFtrends}, we break up the AGN and control samples according to galaxy morphology and stellar mass. The spiral AGN galaxies display a significantly more inside out quenching profile than the spiral control galaxies (Figure~\ref{fig:radialtimeqmorph}). Similarly, the elliptical AGN galaxies have a significantly steeper radial quenching profile than the elliptical control galaxies. Notice that whereas the control quenching profiles differ substantially based on morphology, the AGN quenching profiles are very similar; in both spiral and elliptical AGN galaxies, quenching began in the center $6.5$\,Gyr ago, spreading to the outskirts in $2.5$\,Gyr.

The radial quenching profiles of the AGN and control galaxies are shown by stellar mass in Figure~\ref{fig:radialtimeqSM}. The higher-stellar mass AGN galaxies and the higher-stellar mass control galaxies have similar inside out quenching profiles, whereas lower-stellar mass AGN galaxies differ substantially from lower-stellar mass control galaxies. The differences between the lower-stellar mass AGN and control quenching profiles provides further evidence of AGN feedback taking place in lower-stellar mass galaxies. Similar to morphology, the AGN quenching profiles do not depend strongly on stellar mass; the lower- and higher-stellar mass AGN quenching profiles are similar. The radial quenching trends of the lowest-stellar mass galaxies in our sample ($\log($M$_{\ast}/$M$_\odot)\,{<}\,10.0$) are comparable to the left panel of Figure~\ref{fig:radialtimeqSM}.

The radial quenching profiles demonstrate that the trends seen in the radial SFR profiles of Section~\ref{sec:radialSFRs} are reflective of Gyr-timescale quenching. It is interesting to consider whether the different radial quenching trends of the AGN sample result in different galaxy-wide quenching trends for the subset of currently quiescent AGN galaxies. We explore this possibility below.

\subsection{Galaxy-Wide Quenching Trends}
\label{sec:qtimes}

\begin{figure}
\centering
\includegraphics[width=0.45\textwidth]{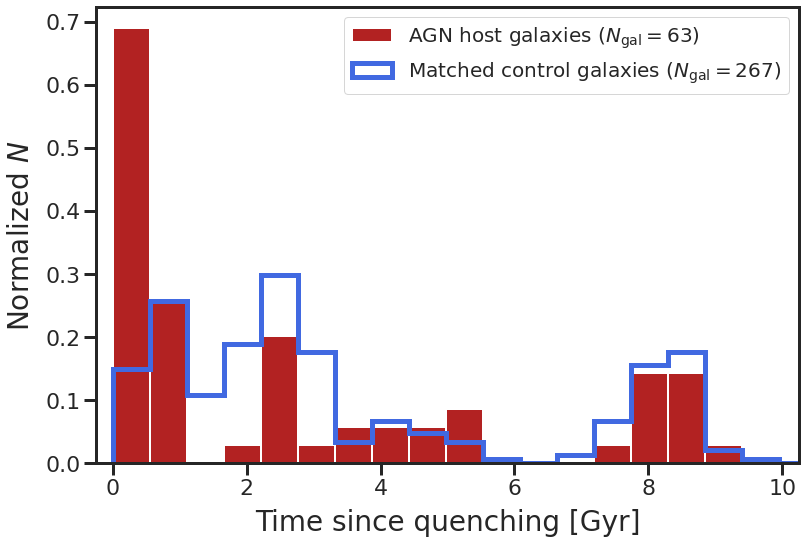}
\caption{Distribution of times since galaxy-wide quenching for the population of quiescent ($\Delta\,\mathrm{SFR}\, < 1.0$\,dex) AGN and control galaxies. When compared with the matched control sample, the AGN sample contains an excess of galaxies whose galaxy-wide SFR quenched recently (in the last $500$\,Myr). The distributions differ at 4.0\,$\sigma$ based on the two-sample KS test (trends persist when excluding ellipticals).}
\label{fig:timeq_hist}
\end{figure}

In all above analyses, we have studied the full AGN and control samples, including both quiescent and nonquiescent galaxies. Here, we explore whether the radial quenching trends of the AGN galaxies lead to different galaxy-wide quenching behavior when compared with the control galaxies. As mentioned in Section~\ref{sec:evolradialSFRs}, the SFMS changes over time, so we cannot simply use a quiescent threshold $1.0$\,dex below the $z\,{\sim}\,0$ PIPE3D SFMS (Figure~\ref{fig:SFMS}). Instead, to quantify the quenching time of the quiescent AGNs, we use the same quenched threshold that we adopted for spaxels: sSFR\,${<}\,0.2/t_{H}(t)$. Figure~\ref{fig:timeq_hist} shows the time since quenching for all quiescent AGN and control galaxies, revealing a clear excess of AGN galaxies that have quenched in the last $500$\,Myr. Typically, this is caused by a recent, short (${\sim}\,1$\,Gyr) quenching period, as illustrated by the example galaxy in Figure~\ref{fig:example}. This is possibly indicative of AGN-driven quenching, but this recent quenching is only seen for a small subset of the AGN galaxies (7\,\%) and does not appear to be tied to the common earlier suppression in central star formation.

In the picture of AGN-driven quenching, the excess of recently quenched galaxies in the AGN sample suggests that AGN activity halts shortly after AGN galaxies quench. In this way, the quiescent AGN sample may consist of galaxies that have quenched recently because they have not yet halted their AGN activity. However, the bulk of the AGN galaxies are not quiescent (Figure~\ref{fig:SFMS}), nor are they preferentially quenching when compared with the control galaxies (Figure~\ref{fig:radialSFRslookback}). Instead, we find that the AGN galaxies reside primarily in the green valley, owing to suppressed central SFRs that arose several gigayears ago.

\section{Discussion}
\label{sec:discussion}

\subsection{AGN Feedback in MaNGA}
\label{sec:AGNquenching}

Spatially resolved SFHs, made possible by the MaNGA survey, allow us to draw conclusions about the detailed effects of AGN feedback on star formation in low-redshift AGNs. Our sample of AGN galaxies largely resides on the SFMS/in the green valley, with far fewer quiescent galaxies than found in the control sample, and a lack of glaring signatures of AGN feedback. However, upon further investigation, we find that the central regions of the AGN galaxies have significantly lower SFRs than the central (kiloparsec-scale) regions of the control galaxies (up to a factor of ${\sim}$\,2). This suppression of central SFRs is responsible for the lack of AGNs on/above the SFMS in our sample, and provides direct observational evidence of AGN feedback suppressing star formation in low-redshift AGNs. Additionally, this suggests that AGN feedback may enhance the central SFR suppression that has been observed in large samples of MaNGA galaxies \citep[e.g.,][]{IbarraPIPE3D2016, BelfioreradialSFR, EllisonradialSFR, Bluckquenching}.

With SFH information, we quantify the timescale over which central quenching occurs in the sample of AGN galaxies. Based on the evolution of radial SFR profiles and radial quenching profiles, a fairly clear picture emerges in which central quenching began in the AGN galaxies ${\sim}\,6$\,Gyr ago ($z\,{\sim}\,0.7$), subsequently taking place over the next few gigayears. As such, the AGN galaxies do not appear to be actively quenching, with their position in the green valley owing to suppression of their central SFRs from several gigayears ago. This is suggestive of a picture in which AGN feedback will not necessarily drive galaxy-wide quenching, possibly due to inefficient coupling to the interstellar medium \citep[ISM; ][]{TorreyFIREAGN}.

Interestingly, Gyr-timescale radial quenching trends in the AGN sample largely persist across spiral/elliptical galaxies and lower-/higher-stellar mass galaxies, whereas the quenching behavior of the control galaxies changes substantially when broken up by morphology and stellar mass. In a small subset of the AGN galaxies, galaxy-wide SFRs were recently quenched (in the last $500$\,Myr) by a short period of quenching (see Figure~\ref{fig:example}), suggesting possible instances of AGN-driven feedback. However, in most cases, some star formation continues in the AGN galaxies at the time of observation, with suppression primarily in the central regions. 

The inside out radial quenching profiles of lower-stellar mass AGN galaxies provide evidence that AGN-driven quenching can also occur in lower-stellar mass galaxies. This contributes to the emerging picture that AGN feedback can suppress star formation in low-stellar mass galaxies \citep[e.g.,][]{PennylowsmAGN, CailowsmAGN, GrahamlowsmAGN, GuolowsmAGN}, contrary to the standard picture in which AGN feedback matters only in the high-stellar mass regime \citep[see also][]{KoudmaniDwarfAGN1, KoudmaniDwarfAGN2}.

At first glance, our results appear to suggest that AGN feedback is typically unsuccessful in quenching low-redshift galaxies. However, the comparable quenching trends of the higher-stellar mass control galaxies to the AGN galaxies raises questions about the quenching mechanism(s) in the control sample. In particular, the population of high-stellar mass, elliptical control galaxies is largely quiescent, with comparable inside out quenching trends to the AGN galaxies. This can be explained by feedback from now-dormant AGNs in the population of control galaxies that were more successful in expelling/heating gas, thereby quenching galaxy-wide SFRs and shutting off AGN activity. This is supported by the finding that there is a missing population of quenched elliptical galaxies in the AGN sample and a buildup of recently quenched AGN galaxies (in the last $500$\,Myr). However, other inside out quenching mechanisms, such as morphological quenching, may also play an important role in quenching high-stellar mass, elliptical galaxies \citep{Martigmorphq, LinMaNGAQuenching, Gensiormorphq}. More work is required to decisively distinguish between AGN feedback and other possible quenching mechanisms in driving the suppression of galaxy-wide SFRs in low-redshift galaxies.

\subsection{Comparison With Other Works}
\label{sec:compare}

Although observations have largely concluded that the gas content of AGN galaxies is comparable to that of non-AGN control galaxies \citep[e.g.,][]{MaiolinoAGNgas, HoAGNgas, VitoAGNgas, ShangguanAGNgas}, it has recently been determined that low-redshift AGNs contain preferentially less gas (factor of ${\sim}$\,2) in their central regions \citep{SanchezAGNSFMS, EllisonAGNgas}. This demonstrates that AGN feedback can deplete host galaxy gas content within the central few kiloparsecs of galaxies, potentially driving star formation quenching. Our results establish that the central regions of AGN host galaxies also have centrally suppressed SFRs (by up to a factor of ${\sim}$\,2), providing direct evidence of AGN feedback suppressing star formation. In a similar vein, it was recently found using ultraviolet imaging that the low-redshift AGN galaxy NGC 3982 has suppressed star formation in its central regions \citep{JosephAGNcentral}. Interestingly, \citet{Woo&EllisonAGNcentral} previously found that the fraction of WISE-identified AGNs in MaNGA is higher among galaxies with enhanced central SFRs. This apparent discrepancy may be caused by the different nature of the studies (``how do AGNs compare with matched control galaxies'' vs.\ ``in which type of galaxy are AGNs found'') or differing AGN selection criteria (optical vs.\ infrared) which can result in significantly different AGN samples \citep{HickoxAGNs, Huang2017AGNs, YaoAGNs}.

Although observations indicate that the central gas depletion timescale is rapid \citep[e.g.,][]{SturmAGNoutflow, CiconeAGNoutflow, FluetschAGNgas}, we find that the suppression of central SFRs arises over several gigayears. This is instead consistent with the picture that integrated AGN accretion is required for AGN feedback to quench star formation, as suggested by the IllustrisTNG simulation \citep{TerrazasAGNTNG, PiotrowskaAGNTNG}. However, whereas the low-accretion rate implementation of AGN feedback in the IllustrisTNG simulation leads to long-term galaxy-wide quenching \citep{WeinbergerAGN, ZingerAGNTNG, NelsonTNGquenching, ParkTNG}, our AGN galaxies are not preferentially in the process of quenching when compared with the control galaxies. The relatively local impact of AGN feedback in our sample of AGN galaxies is instead suggestive of the picture that AGN feedback does not necessarily couple efficiently to the ISM, as found in zoom-in simulations with FIRE \citep{TorreyFIREAGN} and observations of outflows in local AGNs \citep{FluetschAGNgas}. However, the suppressed central SFRs of the AGN galaxies do result in somewhat lower galaxy-wide SFRs, placing the AGNs in the green valley.

\subsection{Caveats}
\label{sec:caveats}

The strength of our conclusions depends on the reliability of our spatially resolved SFHs reconstructed from PIPE3D stellar population information. There are known limitations to the quality of SFH information that can be inferred from optical galaxy spectra \citep{OcvirkSFHlimits, FerrerasSFHlimits}. For instance, SSP age bins that increase with lookback time limit the time resolution of our SFHs; at low lookback times our SFHs are sensitive to Myr-timescale effects, however, at large lookback times these effects would be missed. As such, the agreement between the AGN and control profiles at lookback times of $6$\,Gyr and beyond (Figure~\ref{fig:radialSFRslookback}) indicates Gyr-averaged agreement, with potential deviations on shorter timescales. Beyond these fundamental limitations, SED fitting can also introduce significant biases into the recovered SFHs. The ability of the PIPE3D methodology to recover spatially resolved SFHs has been studied with simulated MaNGA-like data cubes, validating the PIPE3D SED fitting process (\citealt{IbarraMedelMaNGAPIPE3D}; they adopt simple piecewise-constant SFHs rather than nonparametric SFHs). Note, however, that D. Walters et al.\ in preparation questions the reliability of SFH information derived from optical spectra.

To properly interpret our results, it is important to recognize that the control sample consists of several types of ``non-AGN'' galaxies. Firstly, the control sample contains genuine non-AGNs, whose SMBHs have never experienced significant accretion or are too small to enable substantial AGN activity. Secondly, there are galaxies in the control sample that were previously active and have subsequently turned off (e.g., due to a lack of gas). Lastly, the control sample inevitably contains missed AGNs, some of which can be identified by AGN selection criteria at other wavelengths \citep{ComerfordAGNMaNGA}. The representation of these distinct control galaxy types likely depends on galaxy properties (e.g., perhaps more high-stellar mass, quenched control galaxies were previously active), and it is difficult to fully disentangle these populations. As such, comparisons with the matched control sample must be interpreted with the caveat that the control sample is not simply a population of genuine non-AGNs.

\section{Conclusion}
\label{sec:conclusion}

In this work, we have compared the star formation behavior of AGN galaxies and non-AGN control galaxies in the MaNGA survey to study the effects of AGN feedback on low-redshift galaxies. Taking advantage of spatially resolved AGN selection criteria developed for MaNGA (W18), we construct a large sample of 279 optical AGNs. To compare against the AGN sample, we select two redshift- and stellar mass-matched control galaxies for each of the AGN galaxies, resulting in a control sample of 558 galaxies. For straightforward radial stacking and reliable stellar population information, we restrict our attention to face-on galaxies without signatures of type 1 AGN emissions.

With stellar population information from PIPE3D \citep{SanchezMaNGAPIPE3D}, we reconstruct spatially resolved, nonparametric SFHs, following the Gaussian process-based methodology from \citet{Iyerdb2}. We find that our sample of AGN galaxies primarily resides on the SFMS/in the green valley, with fewer galaxies above the SFMS and a missing population of quenched elliptical galaxies. Overall, the galaxy-wide SFRs of the AGNs lack obvious signatures of AGN feedback. However, exploring radial trends in SFR reveals that the central (kiloparsec-scale) SFRs in the AGN galaxies are significantly suppressed when compared with those of the control galaxies (by up to a factor of ${\sim}$\,2). This provides direct evidence of AGN feedback hindering star formation in the central regions of their host galaxies.

With SFH information, we quantify the timescale over which the suppression of central star formation occurs in the AGN galaxies. The evolution of radial SFR profiles, and newly introduced radial quenching profiles, reveal that quenching occurs in the center of the AGN galaxies over Gyr-timescales. Irrespective of morphology and stellar mass, central quenching began in the AGN galaxies ${\sim}$\,$6$\,Gyr ago ($z\,{\sim}\,0.7$), taking place over the next few gigayears. This suggests that integrated AGN accretion is required for AGN feedback to suppress star formation, in agreement with simulations \citep{TerrazasAGNTNG, PiotrowskaAGNTNG}. Furthermore, we find that the AGN galaxies are not preferentially in the process of quenching, with their position in the green valley instead owing to central SFR suppression, which occurred several gigayears ago.

A subset of the AGNs are quiescent at the time of observation, which was, in some cases, caused by a recent, rapid period of quenching that may have been driven by AGN feedback. More frequently, however, the AGN galaxies host some star formation at the time of observation, with centrally suppressed SFRs. This is consistent with the picture that AGN feedback does not necessarily couple efficiently to the ISM, limiting the effectiveness of AGN feedback in driving galaxy-wide quenching (see \citealt{TorreyFIREAGN}). That being said, the similarity in quenching trends between the AGN galaxies and elliptical high-stellar control galaxies suggests that AGN feedback (in no-longer active galaxies) may have played a role in driving these galaxies to quiescence.

This work establishes that AGN feedback can suppress the central star formation of low-redshift galaxies, and provides observational constraints on the extent and timescale of AGN-driven quenching. Our results favor a scenario in which integrated AGN feedback (rather than an instantaneous outflow) is required to significantly affect star formation, and indicate that AGN feedback may be inefficient in driving galaxy-wide quenching in low-redshift galaxies, instead leaving them in the green valley.

\section{Acknowledgments} 
\label{sec:acknowledgments}

We thank the anonymous referee for their careful review and insightful comments. We would also like to thank Sandra Faber, Eric Gawiser, Philip Hopkins, and Rebecca Smethurst for useful feedback and discussions. C.L. and K.G.I. acknowledge the support of the Dunlap Institute for Astronomy \& Astrophysics. Support for K.G.I. was provided by NASA through the NASA Hubble Fellowship grant HST-HF2-51508 awarded by the Space Telescope Science Institute, which is operated by the Association of Universities for Research in Astronomy, Inc., for NASA, under contract NAS5-26555. H.I.M. acknowledges a support grant from the Joint Committee ESO-Government of Chile (ORP 028/2020). S.F.S. thanks the support of the PAPIIT-DGAPA AG100622 project.

Funding for the Sloan Digital Sky Survey IV has been provided by the Alfred P. Sloan Foundation, the U.S. Department of Energy Office of Science, and the Participating Institutions. SDSS-IV acknowledges support and resources from the Center for High-Performance Computing at the University of Utah. The SDSS website is \url{www.sdss.org}. SDSS-IV is managed by the Astrophysical Research Consortium for the Participating Institutions of the SDSS Collaboration including the Brazilian Participation Group, the Carnegie Institution for Science, Carnegie Mellon University, the Chilean Participation Group, the French Participation Group, Harvard-Smithsonian Center for Astrophysics, Instituto de Astrof\'isica de Canarias, The Johns Hopkins University, Kavli Institute for the Physics and Mathematics of the Universe (IPMU)/University of Tokyo, the Korean Participation Group, Lawrence Berkeley National Laboratory, Leibniz Institut f\"ur Astrophysik Potsdam (AIP), Max-Planck-Institut f\"ur Astronomie (MPIA Heidelberg), Max-Planck-Institut f\"ur Astrophysik (MPA Garching), Max-Planck-Institut f\"ur Extraterrestrische Physik (MPE), National Astronomical Observatories of China, New Mexico State University, New York University, University of Notre Dame, Observat\'ario Nacional / MCTI, The Ohio State University, Pennsylvania State University, Shanghai Astronomical Observatory, United Kingdom Participation Group, Universidad Nacional Aut\'onoma de M\'exico, University of Arizona, University of Colorado Boulder, University of Oxford, University of Portsmouth, University of Utah, University of Virginia, University of Washington, University of Wisconsin, Vanderbilt University, and Yale University.

This work relied on the following software: Marvin \citep{CherinkaMarvin}, george \citep{Ambikasarangeorge}, Astropy \citep{Astropyv5.0}, SciPy \citep{VirtanenSciPy}, NumPy \citep{HarrisNumPy}, Matplotlib \citep{HunterMatplotlib}, and seaborn \citep{Waskomseaborn}.

\appendix

\section{More Restrictive AGN/Control Samples}
\label{sec:selection}

\begin{figure*}
\centering
\includegraphics[width=0.9\textwidth]{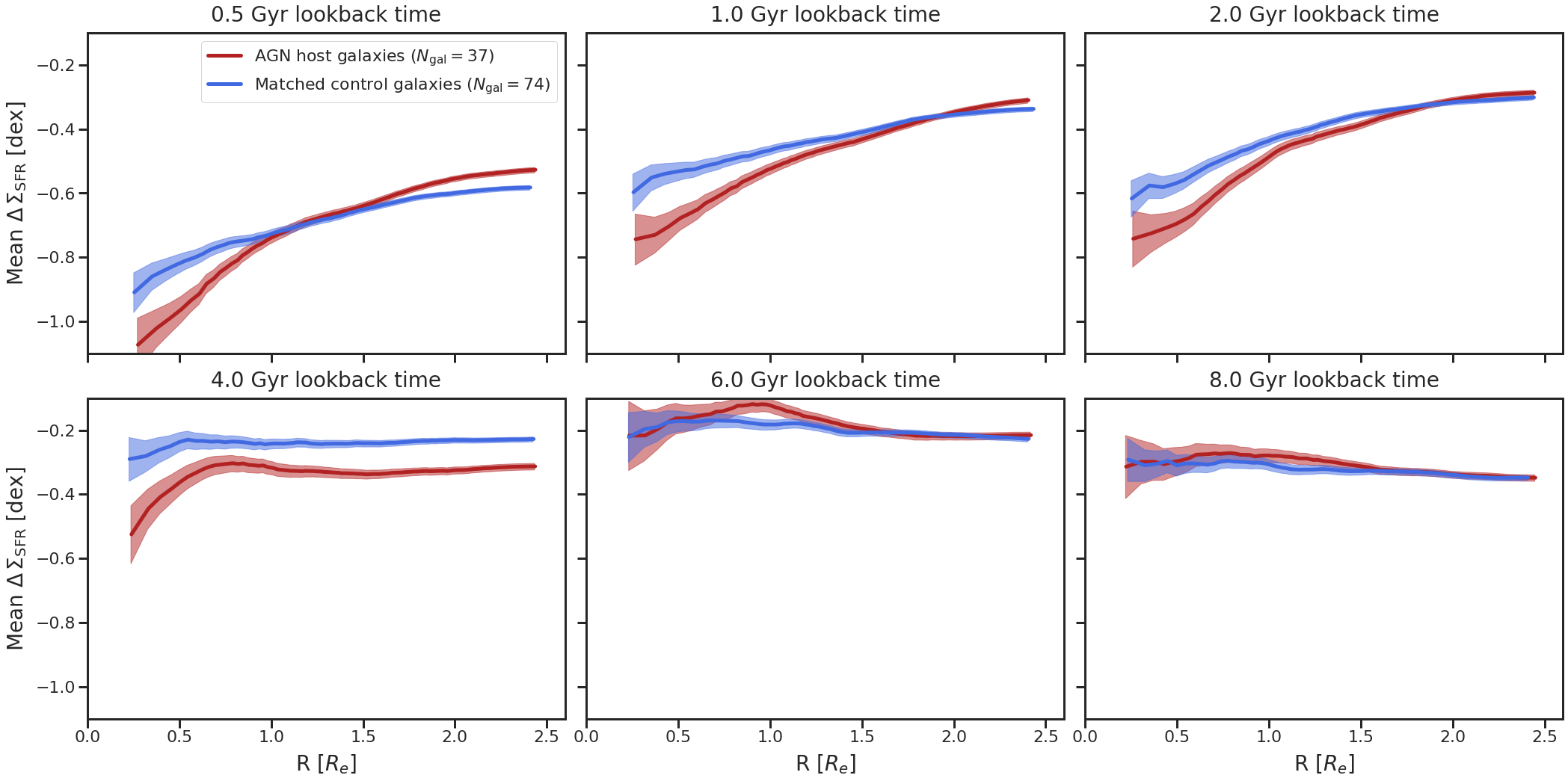}
\caption{Same as Figure~\ref{fig:radialSFRslookback}, repeated on the subset of 37 galaxies in our AGN sample that pass unresolved BPT/WHAN criteria, and control galaxies that are matched in SFR, in addition to stellar mass and redshift. Radial trends for the subset of 37 AGNs closely agree with those of the full 287 AGN sample in Figure~\ref{fig:radialSFRslookback}. Matching the control galaxies in SFR leads to smaller differences between the AGN and control samples, but the AGN galaxies still have statistically significant (based on the $3\,\sigma$ error bars) suppressed central SFR densities, despite the small number of galaxies (arising from tens of thousands of spaxels).}
\label{fig:radialSFRslookbacksub}
\end{figure*}

In Section~\ref{sec:methods}, choices were made while constructing the AGN and control samples that could affect our subsequent results. In particular, to study a large sample of AGN galaxies, we adopted the W18 selection criteria, created for spatially resolved observations. This selection methodology risks the inclusion of some non-AGNs in the AGN sample, misidentified due to other processes that can mimic AGN signatures (e.g., shocks and young hot stars). Previous studies of AGN host galaxies with MaNGA have typically adopted more conventional, unresolved BPT/WHAN selection criteria \citep{first62AGNI, SanchezAGNSFMS}, resulting in smaller AGN samples with (likely) fewer misidentifications. A total of 37 AGNs in our 287 AGN sample overlap with the samples studied in \citet{first62AGNI} and \citet{SanchezAGNSFMS} (some of these galaxies were removed by the edge-on/type 1 AGN cuts, but most were simply not selected; see W18 for discussion on this). When we restrict our analysis to these 37 AGNs that pass the unresolved BPT/WHAN criteria, we obtain very similar results. As an illustration of this, Figure~\ref{fig:radialSFRslookbacksub} shows the evolution of the radial SFR profile for the subset of 37 AGNs; trends closely match those of the full 287 AGN sample in Figure~\ref{fig:radialSFRslookback} (with larger uncertainties).

When selecting the control galaxies, we matched the stellar masses and redshifts of the AGNs, but not the SFRs, which allowed us to study trends in galaxy-wide star formation (Figs.~\ref{fig:SFMS},~\ref{fig:SFRoffsets}). Here, we also compare the radial trends of the AGN galaxies to those of the control galaxies with matched SFRs. For each of the 37 AGNs in Figure~\ref{fig:radialSFRslookbacksub}, we select two control galaxies matched in stellar mass, redshift, and SFR from the 558 control galaxy sample (following the process in Section~\ref{sec:controlsample}). The resulting sample of 74 control galaxies closely matches the stellar masses, redshifts, and SFRs of the 37 AGNs, and the evolution of the resulting radial SFR profile is shown in Figure~\ref{fig:radialSFRslookbacksub}. With control galaxies that are matched in SFR, the differences between the AGN and control profiles are smaller, but the AGN galaxies still show suppression in their central ${\Delta}\,{\Sigma}_{\mathrm{SFR}}$ (by $\sim{0.2}$\,dex at the smallest $R$). This difference remains statistically significant despite the small number of galaxies because it is a trend among the SFR densities of tens of thousands of spaxels.

\bibliography{refs}{}
\bibliographystyle{aasjournal}

\end{document}